\documentclass[prl,twocolumn,aps,showpacs]{revtex4}

\usepackage{graphicx}
\usepackage{bm}
\usepackage{amsmath}

\newcommand{\ket}[1]{\left| #1 \right>} 
\newcommand{\bra}[1]{\left< #1 \right|} 

\begin{document}

\title{Merging Quantum Loop Gases: a Route to Non-Abelian Topological Phases}

\author{Bel\'{e}n Paredes}

\affiliation{Instituto de F\'{i}sica Te\'{o}rica CSIC/UAM \\C/Nicol\'{a}s Cabrera, 13-15
Cantoblanco, 28049 Madrid, Spain}

\date{\today}

\begin{abstract} 
Condensation of quantum loops naturally leads to topological phases with Abelian excitations. Here, I propose that non-Abelian topological phases can arise from {\em merging} two (or several) identical Abelian quantum loop condensates. I define merging through a symmetrization operation, which makes the two loop condensates indistinguishable, inducing the possibility of a topological degeneracy in the space of quasiparticles. To illustrate the construction, the case of two identical toric-code quantum loop gases is considered.
A spin-$1$ model for the two dimensional square lattice is presented for which the resulting merged state is the exact unique ground state.
This Hamiltonian involves four-body interactions between spins located at the same plaquette or vertex, which are quadratic in the spin-$1$ operators. Vertex and plaquette interaction terms are not mutually commuting. The model displays gapped excitations which result from merging anyons in both toric-code copies. These excitations exhibit non-Abelian braiding properties.
\end{abstract}


\maketitle

\section{Introduction}

Topological states represent an exotic organizational form of quantum matter that contradicts the traditional paradigms of condensed matter physics \cite{Wen90, Wen91}. In these states, particles are ordered following a hidden global pattern which is not associated with the breaking of any symmetry and is revealed in the presence of topological quasiparticles obeying exotic statistics. Our understanding of how topological order emerges from the microscopic degrees of freedom of a quantum many-body system is far from complete. Especially intriguing is the possible formation of non-Abelian topological phases \cite{MooreRead91, Wen91_2, Nayak2008, Stern2010}, whose excitations display non-Abelian braiding properties \cite{Stern2008} with potential application for quantum computing \cite{Nayak2008}. Exploring means by which these phases arise, can both sharpen our theoretical picture of their origin and serve as a guide for their realization in physical systems. 

A physical mechanism for the formation of a large class of topological phases has been shown to be the organization of the particles into effective extended objects such as loops or string-nets \cite{Kitaev2003, Freedman2004, Levin2005, Fradkin2005, Freedman2005, Fendley2008, Troyer2008, Fidkowski2009}. At low temperatures these large-scale structures become highly fluctuating and condense. For instance, in two spatial dimensions, topological phases with Abelian excitations can result from condensation of fluctuating closed loops $\ket{\mathcal{L}}$ in the form:

\begin{equation}
\ket{\Phi} \propto 
\sum_{\{\mathcal{L}\}} 
\ket{\mathcal{L}}.
\label{KitaevState}
\end{equation} 
These quantum loop condensates appear in a variety of spin or dimer lattice models, in which the low-energy Hilbert space is spanned by configurations of closed loops \cite{Kitaev2003, Freedman2004, Fendley2008}. A seminal example is Kitaev's toric code model \cite{Kitaev2003}, which describes spins-$\frac{1}{2}$ on the links of a two dimensional square lattice.  The ground state is a sum over all possible configurations in which links with up-spins form a set of closed loops. Such apparently disordered state, exhibits global hidden order and displays fractionalized Abelian excitations localized at the ends of open loops. 

Whereas condensation of loops typically leads to topological phases with Abelian excitations, non-Abelian topological phases can emerge from condensation of more complex objects such as networks of strings \cite{Levin2005,Fidkowski2009}. A prominent example is the string-net model proposed by Levin and Wen for spins-$\frac{1}{2}$ in the honeycomb lattice \cite{Levin2005}, which supports Fibonacci anyons and realizes a fault tolerant quantum computer \cite{Nayak2008}. String-net models have been shown to be equivalent to generalized loop models involving non-orthogonal inner products \cite{Fendley2008}. The non-orthogonality of loop configurations allows to satisfy the intricate topological constraints required for non-Abelian anyons to exist \cite{Fendley2008,Velenich2010}. Other routes towards non-Abelian phases with loop gases, involve either non-local interactions \cite{Troyer2008} or trivalent graphs \cite{Fradkin2005,Velenich2010}.

\begin{figure}[t!]
\includegraphics[width=\linewidth]{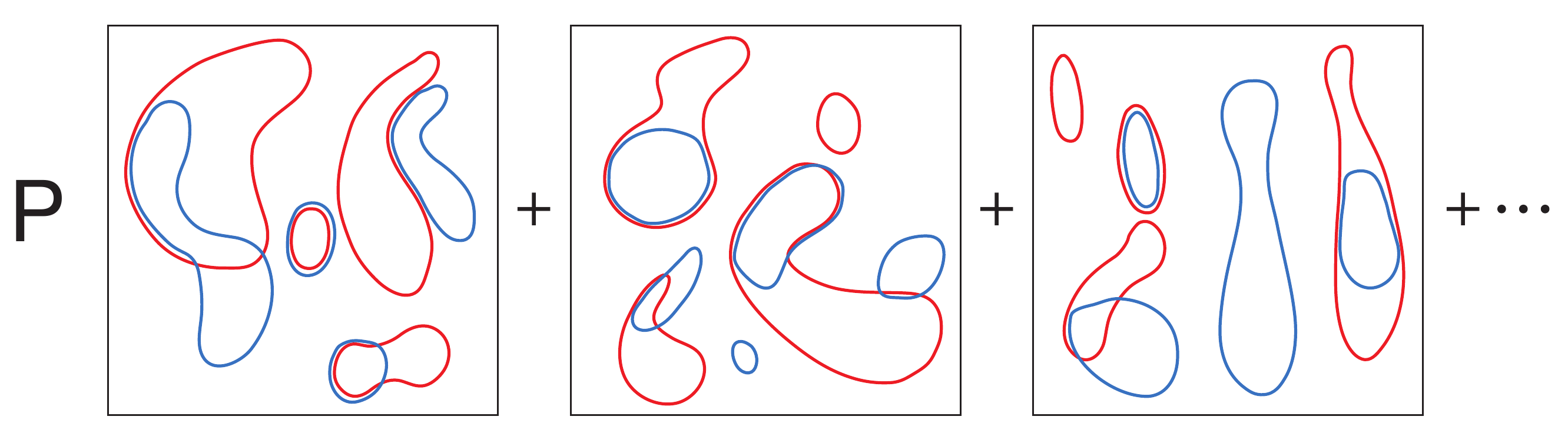}
\caption{\textbf{Merging two quantum loop gases}. Two identical quantum loop gases (red and blue) are merged through the projector $\mathbf{P}$, which unifies the color of lines and converts double lines, where loops coincide, into double-thick lines.
}
\label{MergedLoopGas}
\end{figure}
In this work I propose that non-Abelian topological phases can result from merging two (Abelian) identical quantum loop states in the form:
\begin{equation}
\ket{\Psi}\propto \mathbf{P} \,
\left(
\sum_{\{\mathcal{L}_1\}} 
\ket{\mathcal{L}_1} \,\,\otimes 
\sum_{\{\mathcal{L}_2\}} 
\ket{\mathcal{L}_2}
\right).
\label{MergedState1}
\end{equation}
In this construction, merging is defined by the action of the operator $\mathbf{P}$, which projects the two identical local degrees of freedom $\mathcal{H}\otimes\mathcal{H}$
onto a new degree of freedom $\mathcal{H}_S$ that is symmetric under exchange of the two components. For example, if $\mathcal{H}$ describes a spin $\frac{1}{2}$, then $\mathcal{H}_S$ corresponds to a spin $1$.
The state in Eq.\,(\ref{MergedState1}) involves loop configurations made out of both single and double line segments (Fig.\,\ref{MergedLoopGas}). The distribution of segments seems lawless, but it obeys, by construction, a hidden global rule: each configuration can be decomposed into two closed loop configurations. Besides the condensation of closed loops in each of the copies, this class of states involves a  ''condensation" of the two loop condensates, as if they were indistinguishable macroscopic bosonic objects.
This indistinguishability, achieved by the symmetrizing operator $\mathbf{P}$, makes this family of states potential carriers of non-Abelian excitations. If pairs of anyons are created by cutting loops in each of the identical condensates, a freedom exists to assign the quasiparticles to each of the copies. But states corresponding to different assignments are only globally distinct: pairs belonging to the same copy are attached by segments of loop, whereas pairs of different copies are not. A topological degeneracy arises that opens up the possibility of non-Abelian braiding. 

\begin{figure}[t!]
\includegraphics[width=\linewidth]{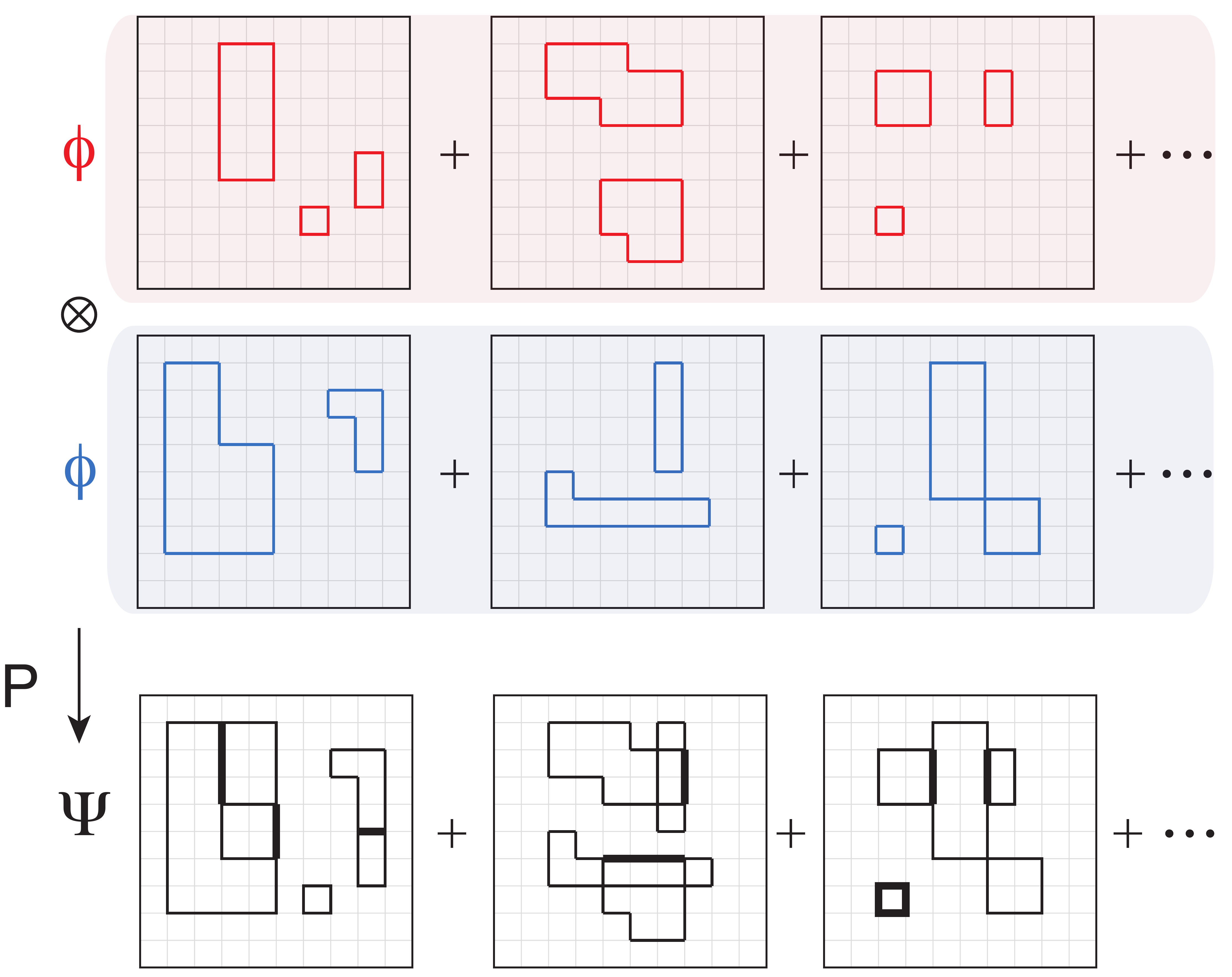}
\caption{\textbf{Merging two toric code ground states}. The toric code ground state, $\Phi$, describes spins $\frac{1}{2}$ sitting at the edges of a square lattice. It is a superposition of closed loop configurations, where each line segment corresponds to a spin up and no-segment corresponds to a spin down. Two copies of the state $\Phi$ are merged by the operator $\mathbf{P}$, leading to the spin-1 state $\Psi$.
}
\label{Spin1Merged}
\end{figure}

To illustrate this construction and to motivate its possible potential for describing non-Abelian phases,  I consider the case of merging two identical toric code ground states (Fig.\,\ref{Spin1Merged}).  The resulting state describes a spin-$1$ system in the square lattice. A local Hamiltonian is constructed which stabilizes this merged quantum loop gas.  Gapped excitations with non-Abelian braiding properties are exactly obtained as the result of merging toric code excitations.

The symmetrization construction I present is inspired by the form of the wave function describing non-Abelian fractional quantum Hall liquids
\cite{MooreRead91}, which can be built from copies of Abelian quantum Hall states \cite{Wen99, WenProjectiveConstruction}.
 
\begin{figure}[t!]
\includegraphics[width=\linewidth]{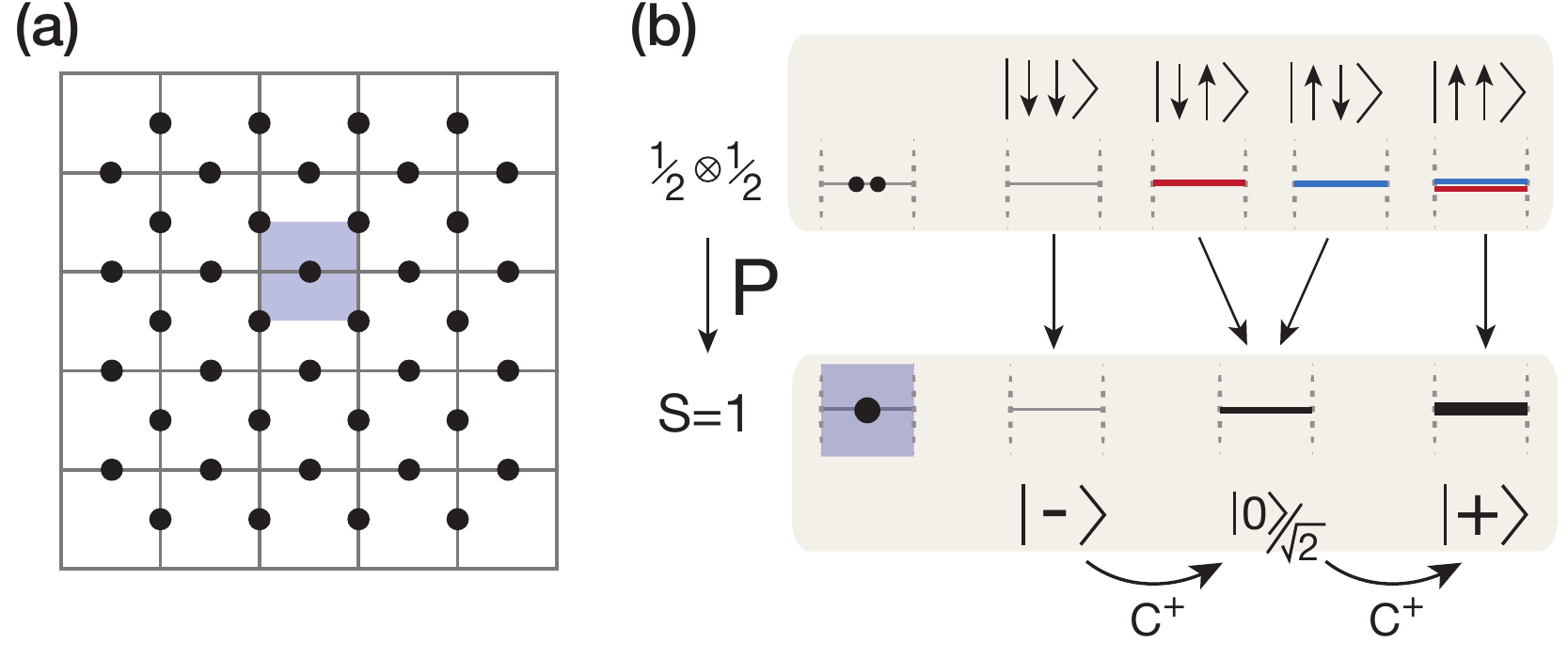}
\caption{\textbf{Spin 1 lattice system}.  
\textbf{a,} Spins-$1$ sit at the edges of a two-dimensional square lattice. \textbf{b,} Projection of two spin-$\frac{1}{2}$ onto a spin-$1$. Spin-$\frac{1}{2}$ and spin-$1$ states are represented by line-segments. For a spin-$1$, the operator $C^\dagger$, defined in the main text, converts a no-line into a single-line, and a single-line into a double-line.
}
\label{Spin1Lattice}
\end{figure}

\section{Spin-1 Merged State}

The toric code model describes spins $\frac{1}{2}$ sitting at the edges of a square two-dimensional lattice.  
The Hamiltonian, $H=(1-A_v)+(1-B_p)$, is a sum of mutually commuting stabilizer operators:
\begin{eqnarray}
A_v=\prod_{\ell \in v} \sigma_{\ell}^{z}, &&
B_p=\prod_{\ell \in p} \sigma_{\ell}^x,
\label{Stabilizers}
\end{eqnarray}
which describe four-body interactions between spins located at the same vertex ($v$) or plaquette ($p$).
The ground state $\ket{\Phi}$ is a quantum loop condensate of the form of Eq.\ref{KitaevState}, where closed loop states are defined as:
\begin{equation}
\ket{\mathcal{L}} = \prod_{\ell \in \mathcal{L}} 
\sigma_{\ell}^x 
\ket{\textrm{vac}}. 
\label{ClosedLoop}
\end{equation}
Here, $\ket{\textrm{vac}} =\bigotimes_{\ell=1}^N\ket{\downarrow}_\ell$, $\sigma_{\ell}^\alpha$, $\alpha=x,y,z$ is the Pauli matrix at site $\ell$, and $\mathcal{L}$ is a set of closed paths in the square lattice.
The state $\ket{\Phi}$ is invariant under the stabilizers in Eq.\,(\ref{Stabilizers}). 
A crucial property of the state $\ket{\Phi}$ is its self duality \cite{Kitaev2003,Fendley2008}: it is also described as a quantum loop condensate in the dual lattice, for which the role of vertices and plaquettes is exchanged. If $\mathsf{H}$ is the Hadamard unitary transformation mapping the $z$-basis of a spin $\frac{1}{2}$ onto the $x$-basis, we have that 
\begin{eqnarray}
\ket{\Phi}=\mathsf{H}^{\otimes N}\widetilde{\ket{\Phi}}, 
&& \widetilde{\ket{\Phi}} \propto	\sum_{\{\widetilde{\mathcal{L}}\}} 
	\widetilde{\ket{\mathcal{L}}},
\label{KitaevDual}
\end{eqnarray}
where $\widetilde{\mathcal{L}}$ is a set of closed loops in the dual lattice.

Let me consider the state in Eq.\,(\ref{MergedState1}), for the case in which two copies of the toric ground state are merged.
The projector that carries out the symmetrization is  
$\mathbf{P}=\prod_\ell P_\ell$, 
where $P_\ell^{}=\mathbf{1}_\ell^{}-\ket{\phi}_\ell \bra{\phi}_\ell$, 
and $\ket{\phi}_\ell=(\ket{\uparrow \downarrow}_\ell-\ket{\downarrow \uparrow}_\ell)/\sqrt{2}$ 
is the singlet state.  The resulting state describes spins $1$ in the square lattice in the form:
\begin{equation}
\ket{\Psi} \propto 
\sum_{\{\mathcal{L}_1\}  \{\mathcal{L}_2\}} 
\prod_{\begin{array}{c} {\scriptstyle \ell_1 \in \mathcal{L}_1} \\ {\scriptstyle \ell_2 \in \mathcal{L}_2} \end{array}}
\!C^\dagger_{\ell_1}C^\dagger_{\ell_2}
\ket{-},
\label{Spin-1State}
\end{equation}
where $\ket{-}=\bigotimes_{\ell=1}^N\ket{-}_\ell$, and $\ket{-}_\ell$ denotes the  state with minimum $z$ component of the $\ell$th spin-$1$. The operators $C^\dagger_\ell$ correspond to projections of tensor products of spin-$\frac{1}{2}$ operators onto the symmetric subspace in the form:
\begin{eqnarray}
&C^\dagger_\ell\ket{-}  &\equiv  
P_\ell^{} (\sigma^x_\ell \otimes \mathbf{1}^{}_\ell) \ket{\textrm{vac}}=P_\ell^{} (\mathbf{1}_\ell^{}\otimes \sigma^x_\ell) \ket{\textrm{vac}}, \nonumber\\
&[C^\dagger_\ell]^2 \ket{-} &\equiv  P_\ell^{} (\sigma^x_\ell \otimes \sigma^x_\ell) \ket{\textrm{vac}}.
\label{Spin-1Operator4}
\end{eqnarray}
They can be written as $C^\dagger_\ell=\frac{1+S^z_\ell}{2}S^+_\ell$, where $S^+_\ell=S^x_\ell+iS^y_\ell$, $S^\alpha_\ell$, $\alpha=x,y,z$ are spin-$1$ operators. The three orthogonal states of the local spin-$1$ degree of freedom,
$\ket{-}_\ell$, 
$C^\dagger_\ell\ket{-}_\ell=\ket{0}_\ell/\sqrt{2}$, 
and $[C^\dagger]^2_\ell\ket{-}_\ell=\ket{+}_\ell$ are mapped, respectively, onto no-segment, single-line segment, and double-line segment in the corresponding loop configuration (Fig.\,\ref{Spin1Lattice}).

\begin{figure}[t!]
\includegraphics[width=\linewidth]{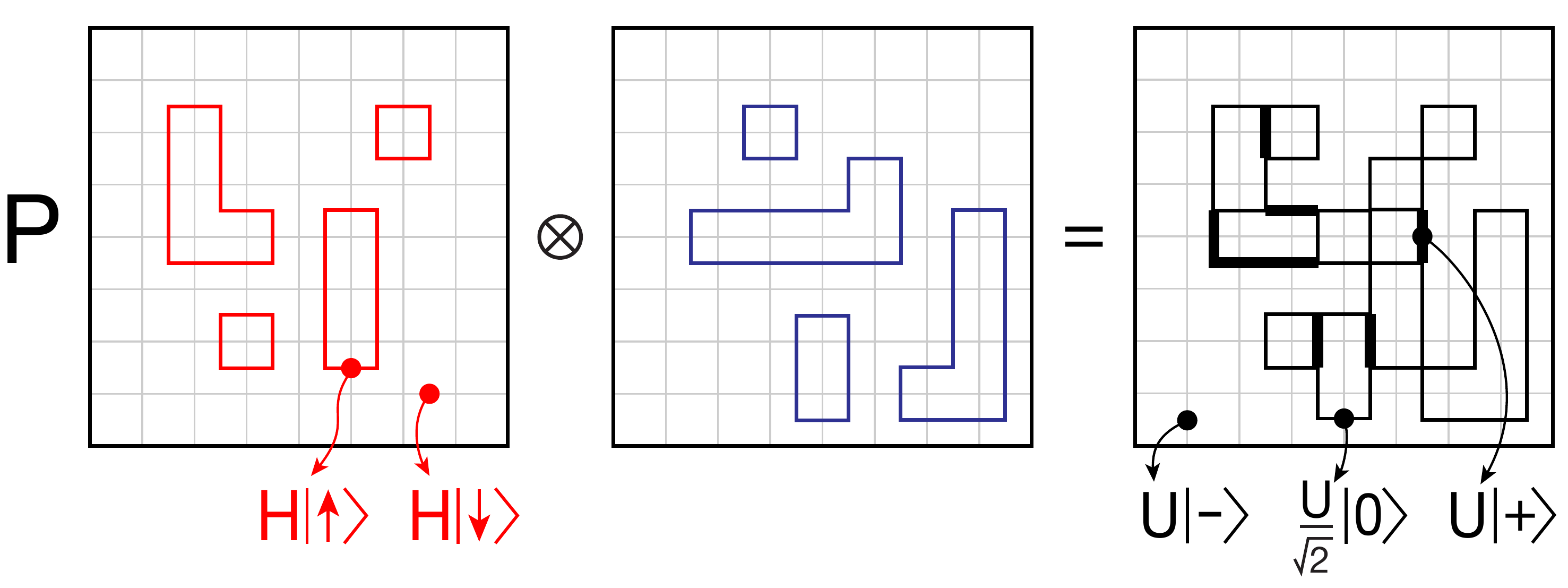}
\center
\caption{\textbf{Self-dual property}. In the dual lattice, a toric-code loop condensate is also a superposition of closed loop configurations, with line segments in the dual lattice related to those in the lattice by the Hadamard transformation $\mathsf{H}$. The spin-$1$ merged state has also the same form in the lattice and in the dual lattice, with dual line-segments related to line-segments by the unitary transformation $U$.}
\label{Duality}
\end{figure}
As a consequence of the self duality of the toric-code components, the state in Eq.\,\ref{Spin-1State} has the same form in the dual lattice
(Fig.\,\ref{Duality}):
\begin{eqnarray}
	\ket{\Psi}=U^{\otimes N}\widetilde{\ket{\Psi}},
	&& U=P_\ell(\mathsf{H}\otimes\mathsf{H})P_\ell,
\label{DualProperty}
\end{eqnarray}
where $\widetilde{\ket{\Psi}}\propto \mathbf{P}  (\widetilde{\ket{\Psi}}\otimes \widetilde{\ket{\Psi}})$. Here, we have used the fact that $\mathsf{H}\otimes \mathsf{H}$ is symmetric under exchange of the two components and therefore commutes with the projector $P_\ell$. The unitary transformation $U=e^{i\frac{\pi}{2}S^y_\ell}$ exchanges the role of $x$ and $z$ in the spin-1 operators, 
$U S^z_\ell U^\dagger=S^x_\ell$. Exploiting the self dual property of the merged state $\ket{\Psi}$ we can construct a Hamiltonian that stabilizes it.

\section{Spin-1 Hamiltonian}

\begin{figure}[t!]
\includegraphics[width=\linewidth]{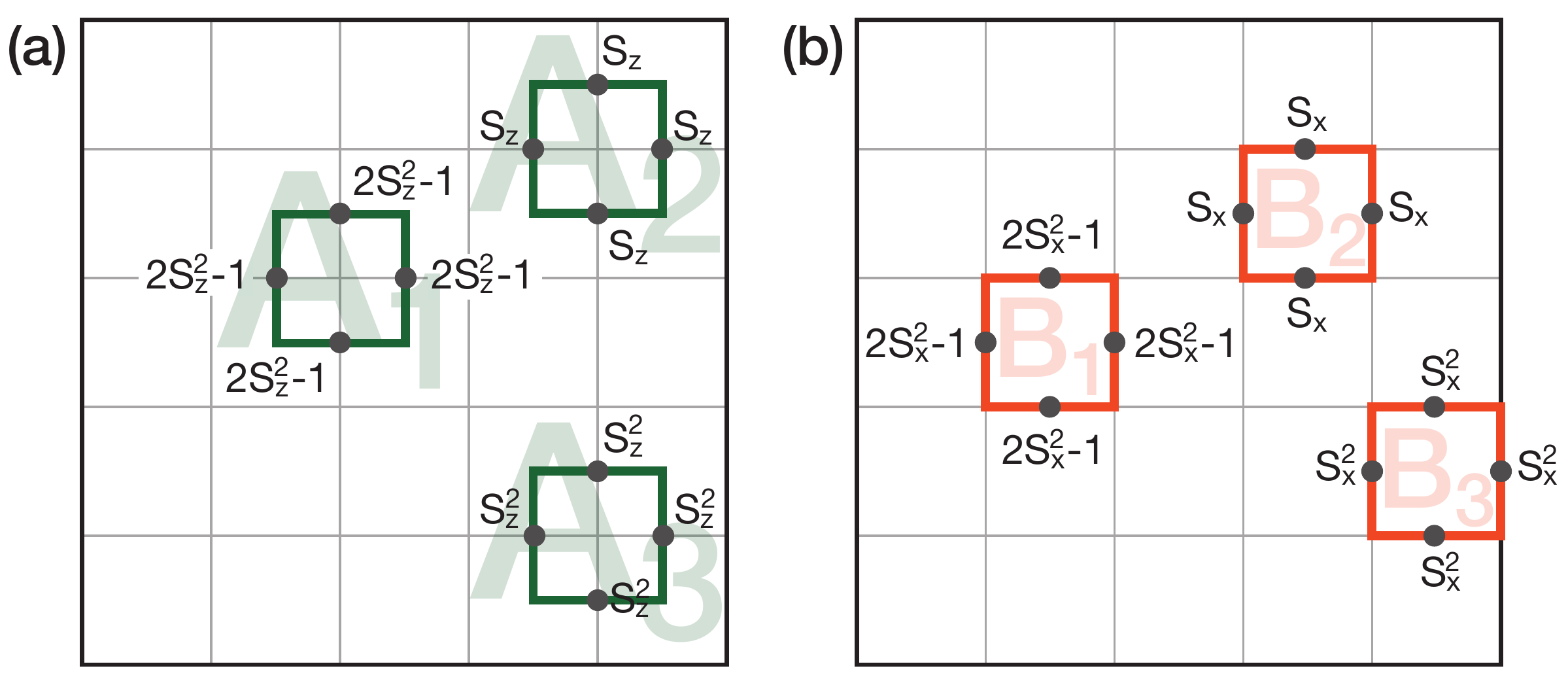}
\center
\caption{\textbf{Vertex and plaquette operators}.
\textbf{a,} Vertex operators $A_{1v}$, $A_{2v}$, $A_{3v}=A_{2v}^2$, and   \textbf{b,} plaquette operators $B_{1p}$, $B_{2p}$, $B_{3p}=B_{2p}^2$ participating in the spin-1 Hamiltonian. Vertex and plaquette operators are not always mutually commuting. We have $\left[A_{1v},B_{ip}\right]=\left[B_{1p},A_{iv}\right]=0$, but $\left[A_{2v},B_{2p}\right]\neq 0$.
}
\label{VertexPlaquetteFig}
\end{figure}

Let me introduce the spin-1 Hamiltonian:
\begin{eqnarray}
H=J\left[\sum_{\ell \in v}(Q_{1v}+Q_{2v})+\sum_{\ell \in p}(Q_{1p}+Q_{2p})\right], 
\label{Hamiltonian}
\end{eqnarray}
where $J$ is a positive constant and $Q_{1v(p)}$,  $Q_{2v(p)}$ are two type of vertex (plaquette) projectors of the form:
\begin{eqnarray}
Q_{1v}=(1-A_{1v})/2 && Q_{2v}=A_{2v}(A_{2v}-1)/2\nonumber \\
Q_{1p}=(1-B_{1p})/2 && Q_{2p}=B_{2p}(B_{2p}-1)/2.
\label{Hamiltonian2}
\end{eqnarray}
These projectors are defined through two kind of vertex and plaquette operators (Fig.\,\ref{VertexPlaquetteFig}):
\begin{eqnarray}
&&A_{1v} =\prod_{\ell \in v} (2[S_{\ell}^z]^2-1), \,\,A_{2v} =\prod_{\ell \in v} S_{\ell}^z\\
&&B_{1p} =\prod_{\ell \in p} (2[S_{\ell}^x]^2-1), \,\,B_{2p} =\prod_{\ell \in p} S_{\ell}^x.
\end{eqnarray}
As in Kitaev's model, plaquette operators correspond to vertex operators in the dual lattice,
$B_{i p}=U^{\otimes N} A_{i \widetilde{v}} U^{\dagger\otimes N}$, with $i=1,2$, and $\widetilde{v}$ denoting a vertex in the dual lattice corresponding to the plaquette $p$. Plaquette projectors, $Q_{ip}$, and vertex projectors, $Q_{iv}$, are therefore dual to each other. 
Importantly, in contrast to what happens in Kitaev's model, plaquette and vertex operators are not always mutually commuting. We have:
\begin{eqnarray}
	\left[Q_{1v},Q_{ip}\right]=0, \,\,\,\left[Q_{1p},Q_{iv}\right]=0, \,\,\, \left[Q_{2v},Q_{2p}\right]\neq 0.
\end{eqnarray}
Finding the eigenstates of the Hamiltonian in Eq.\,(\ref{Hamiltonian}) is therefore not straightforward. 

\section{Ground State}
\begin{figure}[h]
\includegraphics[width=\linewidth]{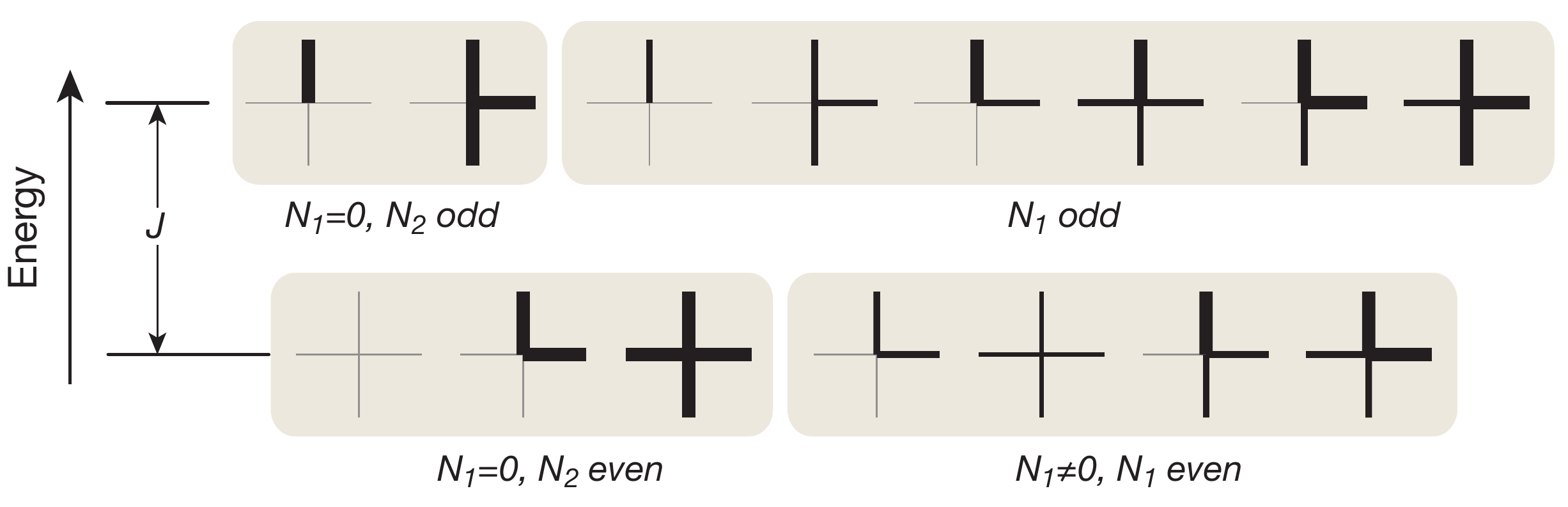}
\center
\caption{\textbf{Zero-energy and excited vertex configurations}. Zero-energy vertex configurations are characterized by two properties: 1) an even number of single-lines, $N_1$, and 2) an even number of double-lines, $N_2$, if there are no single-lines present. Excited vertices are therefore characterized by either having an odd number of single-lines or having an odd number of double-lines with no-single lines present. }
\label{VertexGroundstates}
\end{figure}

The zero-energy ground state subspace is the subspace of states simultaneously annihilated by all projectors $Q_{iv(p)}$. It is thus characterized by the set of conditions:
\begin{eqnarray}
\label{VertexConditions} A_{1v} \ket{\Psi}=\ket{\Psi}, && A_{2v}(A_{2v}-1)\ket{\Psi}=0 \\
\label{PlaquetteConditions} B_{1p} \ket{\Psi}=\ket{\Psi}, && B_{2p}(B_{2p}-1)\ket{\Psi}=0.
\end{eqnarray}
The first vertex condition in Eq.\,(\ref{VertexConditions})  selects vertices with an even number of single lines, $N_{1v}=\sum_{\ell \in v}\ket{0}_\ell\!\bra{0}_\ell$. 
The second vertex condition states that if no single-lines are present at a vertex, the number of double-lines, 
$N_{2v}=\sum_{\ell \in v}\ket{+}_\ell\!\bra{+}_\ell$, must be even.  Together, these conditions define the subspace of vertices indicated 
in Fig.\,\ref{VertexGroundstates}. Similarly, the plaquette conditions in Eq.\,(\ref{PlaquetteConditions}) select the same type of vertex configurations in the dual lattice.

The first plaquette condition in Eq.\,(\ref{PlaquetteConditions}) states that the ground state is invariant under the plaquette-move $B_{1p}$.
This plaquette operator can be written as $B_{1p} = \prod_{\ell \in p} X_\ell$, with 
$X_\ell=\ket{+}_\ell\!\bra{-}_\ell+\ket{-}_\ell\!\bra{+}_\ell$. Within a plaquette, it converts double-lines into no-lines and vice-versa. This plaquette-move leaves the subspace of states satisfying the vertex conditions in Eq.\,(\ref{VertexConditions}) invariant. The plaquette-moves  $B_{2p}$ and $B_{3p}=B_{2p}^2$ can, however, modify the parity of double-lines coinciding at a vertex, creating excited vertices that violate the second condition in Eq.\,(\ref{VertexConditions}). The second plaquette condition in Eq.\,(\ref{PlaquetteConditions}) states that $B_{2p}$-moves are always cancelled by $B_{3p}$-moves.
\newline

Let me show that the merged state $\ket{\Psi}$ in Eq.\,(\ref{Spin-1State}) is an exact ground state of the Hamiltonian in Eq.\,(\ref{Hamiltonian}). This property follows from the fact that the merged components $\ket{\Phi}$ are ground states of the toric code model. Let me start by proving that the state $\ket{\Psi}$ fulfills the vertex conditions in Eq.\,(\ref{VertexConditions}). 
The vertex operator $A_{1v}$ can be written as
\begin{eqnarray}
A_{1v}=\prod_{\ell \in v} P_\ell^{} (\sigma^z_\ell \otimes \sigma^z_\ell)P_\ell= \mathbf{P}(A_{v}\otimes A_{v})\mathbf{P}.
\label{A1v}
\end{eqnarray}
Taking into account that $A_{v}\ket{\Phi}=\ket{\Phi}$, and that $A_{v}\otimes A_{v}$ is symmetric under exchange of the two components and therefore commutes with the projection $\mathbf{P}$, we have that
%
\begin{eqnarray}
A_{1v} \ket{\Psi}=\mathbf{P}(A_{v}\otimes A_{v})\ket{\Phi}\otimes\ket{\Phi}=\ket{\Psi}.
\label{Parity1Even}
\end{eqnarray}
Similarly, we have that  
%
\begin{eqnarray}
&&[A_{2v}]^2 \ket{\Psi} =	\mathbf{P}\prod_{\ell \in v} 
	\frac{\mathbf{1}_\ell^{}\otimes \mathbf{1}_\ell^{} +\sigma^z_\ell \otimes \sigma^z_\ell}{2}\ket{\Psi}=\nonumber\\
&&=\mathbf{P} \prod_{\ell \in v} 
\frac{\sigma^z_\ell \otimes \mathbf{1}_\ell^{}+\mathbf{1}_\ell^{}\otimes \sigma^z_\ell}{2} (A_v \otimes \mathbf{1}_v) 
\ket{\Phi}\otimes\ket{\Phi}=\nonumber\\
&&=A_{2v}\ket{\Psi}.
\end{eqnarray}
The plaquette conditions in Eq.\,(\ref{PlaquetteConditions}) follow directly from the vertex conditions above, as a result of the dual property of the merged state $\ket{\Psi}$ expressed in Eq.\,(\ref{DualProperty}). Since $Q_{i\tilde{v}}\widetilde{\ket{\Psi}}=0$, we have that
\begin{eqnarray}
Q_{ip}\ket{\Psi}=U^{\otimes N} Q_{i\tilde{v}}U^{\dagger \otimes N} U^{\otimes N}\widetilde{\ket{\Psi}}=0.
\end{eqnarray}
It is illuminating to explicitly see how excited vertex configurations created by plaquette-moves $B_{2p}$ and $B_{3p}$ cancel each other in the ground state superposition, so that the second plaquette condition in Eq.\,(\ref{PlaquetteConditions}) is fulfilled. This indeed provides a proof of the uniqueness of the ground state.

\subsection{Uniqueness of the ground state}
The state $\ket{\Psi}$ is the unique ground state of the Hamiltonian in Eq.\,(\ref{Hamiltonian}) on the sphere. This can be seen through the following steps. Let me start by characterizing the subspace of states satisfying the vertex conditions in Eq.\,(\ref{VertexConditions}). This subspace is spanned by graphs in which single-line closed loops are either joined or split by double-line paths. There are two type of graphs of this kind: 1) those that can be decomposed into two single-line closed loop configurations  (Fig.\,\ref{TwoClassGraphs}a), and 2) those that can not (Fig.\,\ref{TwoClassGraphs}b). These two type of graphs are topologically different. The first type can be converted into a closed loop configuration, with no double lines present, by performing a sequence of plaquette moves $B_{1p}$. This is not possible, however, for the second type. The second class of graphs can not participate in the ground state (see Appendix), since the second plaquette condition in Eq.\,(\ref{PlaquetteConditions})  could not be fulfilled in such case. We are then left with the subspace of states of the form:
\begin{figure}[t!]
\includegraphics[width=\linewidth]{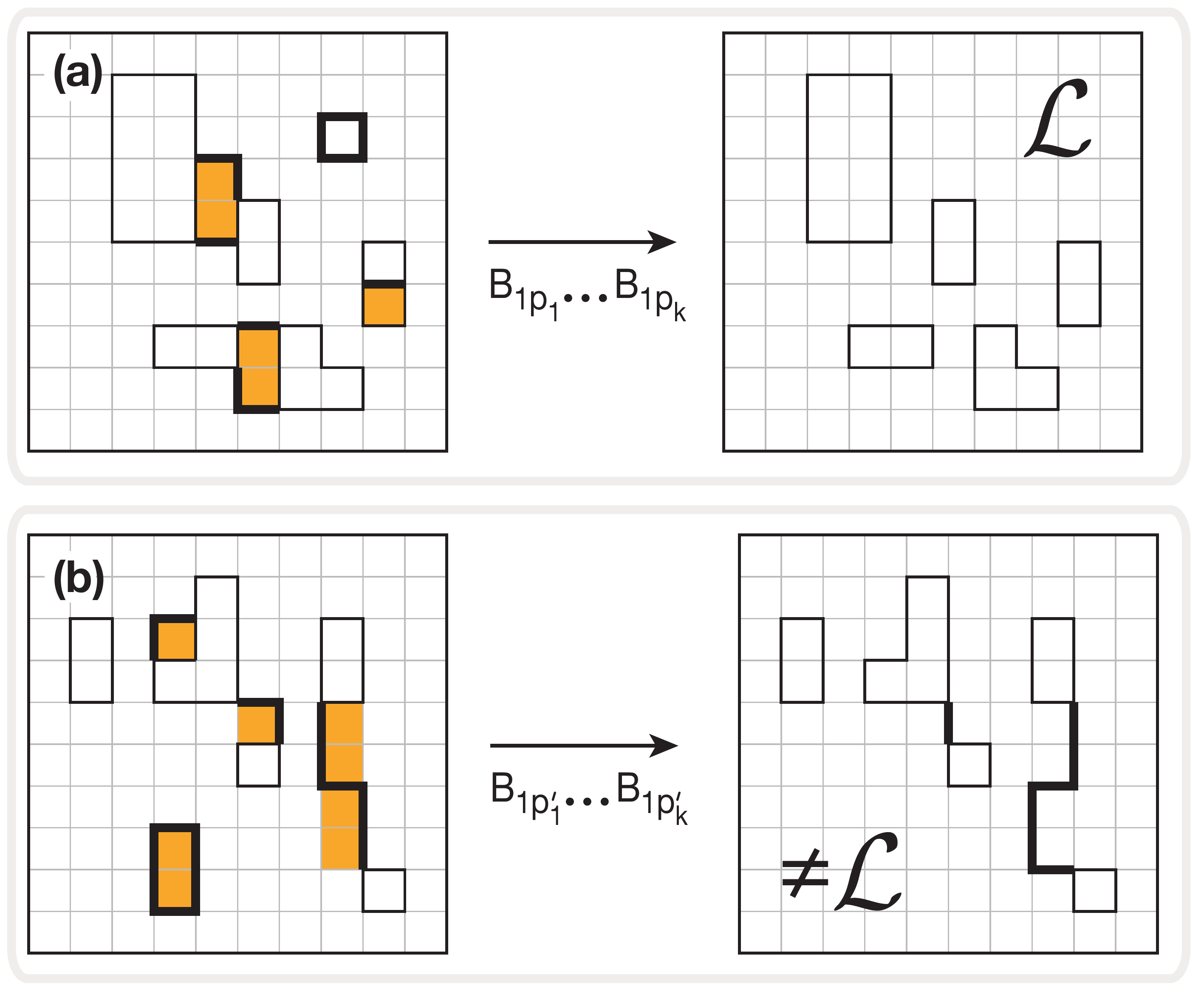}
\center
\caption{\textbf{Zero-energy eigenstates of the vertex Hamiltonian}. The subspace of states fulfilling the vertex conditions is spanned by two classes of configurations: \textbf{a,} those that can be decomposed into two sets of closed paths and \textbf{b,} those that can not. States of the first class are equivalent through a sequence of plaquette moves $B_{1p}$ to a set of single-line closed paths, whereas states of the second class are not.}
\label{TwoClassGraphs}
\end{figure}
\begin{eqnarray}
B_{1p_1} \ldots B_{1p_k}\ket{\mathcal{L}_s},
\label{PlaquetteMoves} 
\end{eqnarray}
where
\begin{eqnarray}
\ket{\mathcal{L}_s}=	\prod_{\ell \in \mathcal{L}} C^\dagger_{\ell}\ket{-}
\end{eqnarray}
is a spin-1 single-line closed loop configuration. The states in Eq.\,(\ref{PlaquetteMoves}) span the merged ground-state in Eq.\,(\ref{Spin-1State}).

The first plaquette condition, $B_{1p}\ket{\Psi}=\ket{\Psi}$, implies that all states that are equivalent (through a sequence of $B_{1p}$ moves) to the same $\ket{\mathcal{L}_s}$ must have equal coefficients, and that therefore, the ground state is of the form
\begin{eqnarray}
\ket{\Psi} \propto	\prod_p \left(1+B_{1p}\right) \sum_{\{\mathcal{L}\}}\beta_\mathcal{L} \ket{\mathcal{L}_s}.
\label{GenGroundState}
\end{eqnarray}
Finally, the second plaquette condition,
\begin{eqnarray}
\langle \mathcal{L}_s\vert B_{2p}\vert \Psi \rangle =\langle \mathcal{L}_s\vert B_{3p} \vert \Psi \rangle,
\end{eqnarray}
uniquely determines all relative coefficients to be
\begin{eqnarray}
\beta_\mathcal{L}=2^{n_{\mathcal{L}}},
\end{eqnarray}
where $n_{\mathcal{L}}$ is the number of loops in the configuration $\mathcal{L}$ (see Appendix). For these coefficients, the state in Eq.\,(\ref{GenGroundState}) is precisely the merged state in Eq.\,(\ref{Spin-1State}). Since single-line loop configurations can arise from either the blue or the red copy, two configurations differing by one loop have a relative amplitude $2$.

\subsection{Ground state degeneracy on a surface with non-trivial topology}
Kitaev's model in the annular geometry exhibits two degenerate ground states, $\ket{\Phi}$ and $\mathcal{T}\ket{\Phi}$, which differ in the presence or not of a non-contractible loop $\mathcal{C}$, with
\begin{eqnarray}
\mathcal{T}=	\prod_{\ell \in \mathcal{C}} \sigma_{\ell}^x.		
\end{eqnarray}
They are eigenstates with eigenvalue $+1$ and $-1$, respectively, of the global operator
$\mathcal{F}= \prod_{\ell \in c} \sigma_{\ell}^z$,
where $c$ is a cross cut in the annulus.
\begin{figure}[t!]
\includegraphics[width=\linewidth]{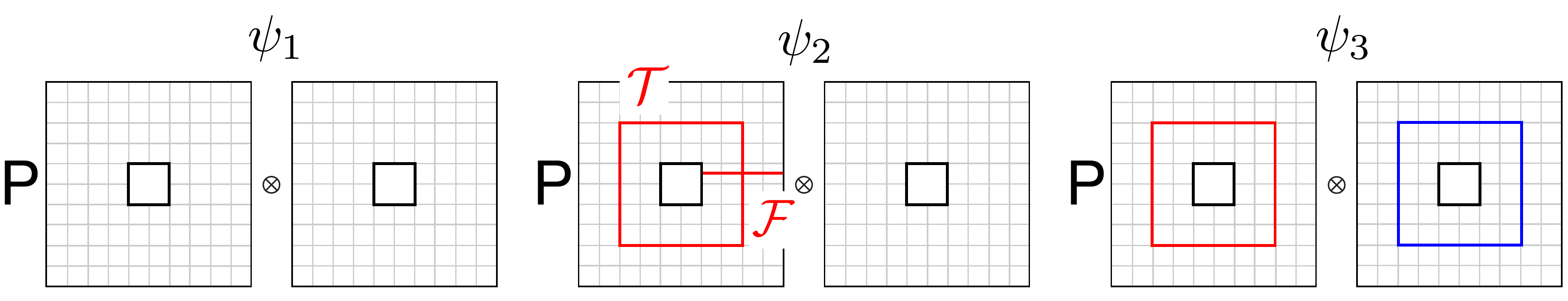}
\caption{\textbf{Ground state degeneracy on the annulus}. 
The spin-1 model is 3-fold degenerate on the annulus. The different states result from the distinct ways of merging the two possible Kitaev ground states on the annulus. The latter are characterized by the presence or not of a non-contractible loop created by the operator $\mathcal{T}$. The presence of such loop is detected by the dual operator $\mathcal{F}$.
}
\label{DegAnnulus}
\end{figure}

\begin{figure}[t!]
\includegraphics[width=\linewidth]{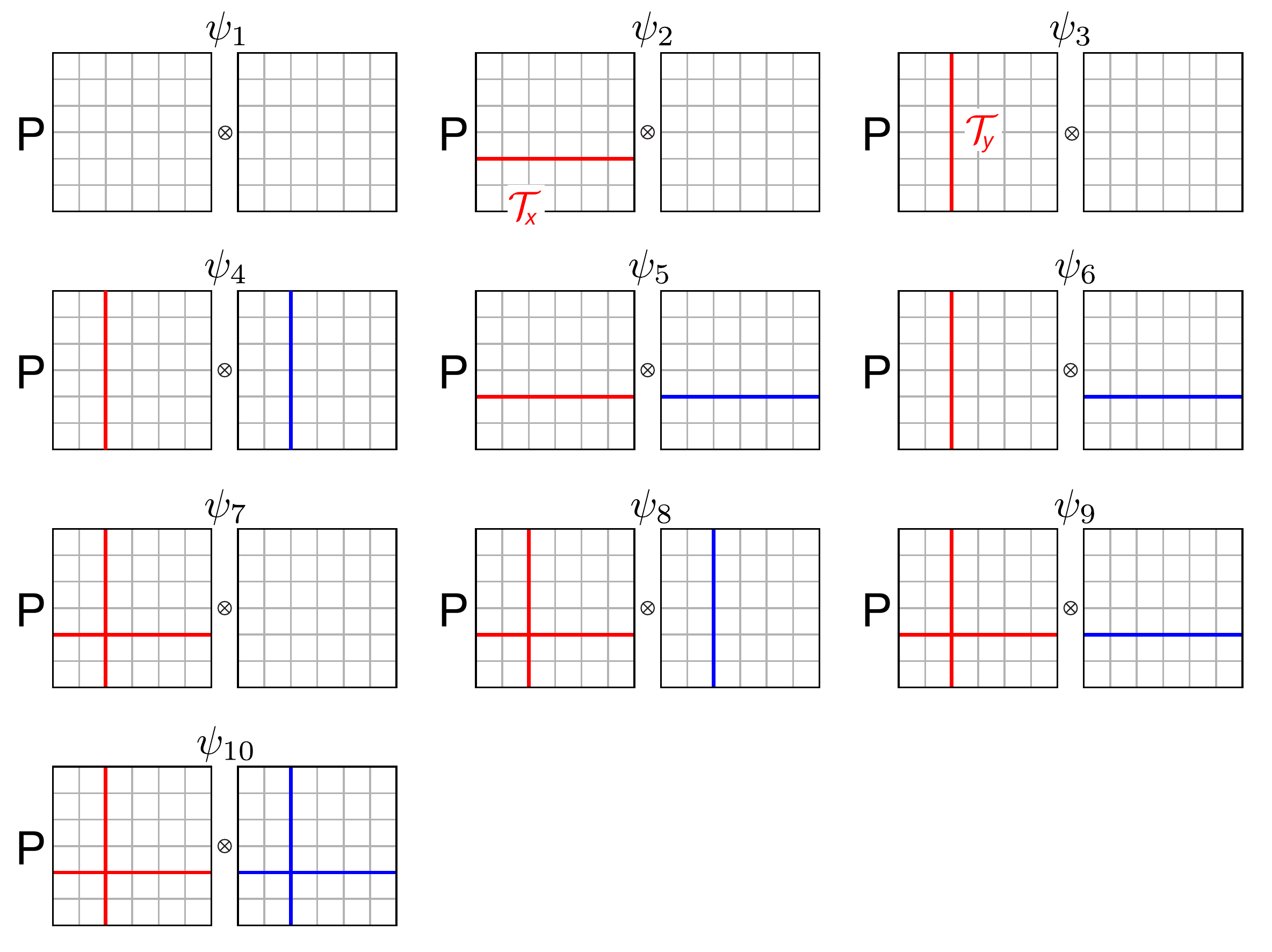}
\center
\caption{\textbf{Ground state degeneracy on the torus}. 
The spin-1 model is 10-fold degenerate on the torus. The different ground states correspond to the distinct states that result from merging the 4 possible Kitaev ground states on the torus. The 4 Kitaev ground states are characterized by the presence or not of one or both the two type of non-contratible loops on the torus, created, respectively, by the operators $\mathcal{T}_x$ and $\mathcal{T}_y$.
}
\label{DegTorus}
\end{figure}

\begin{figure}[t!]
\includegraphics[width=\linewidth]{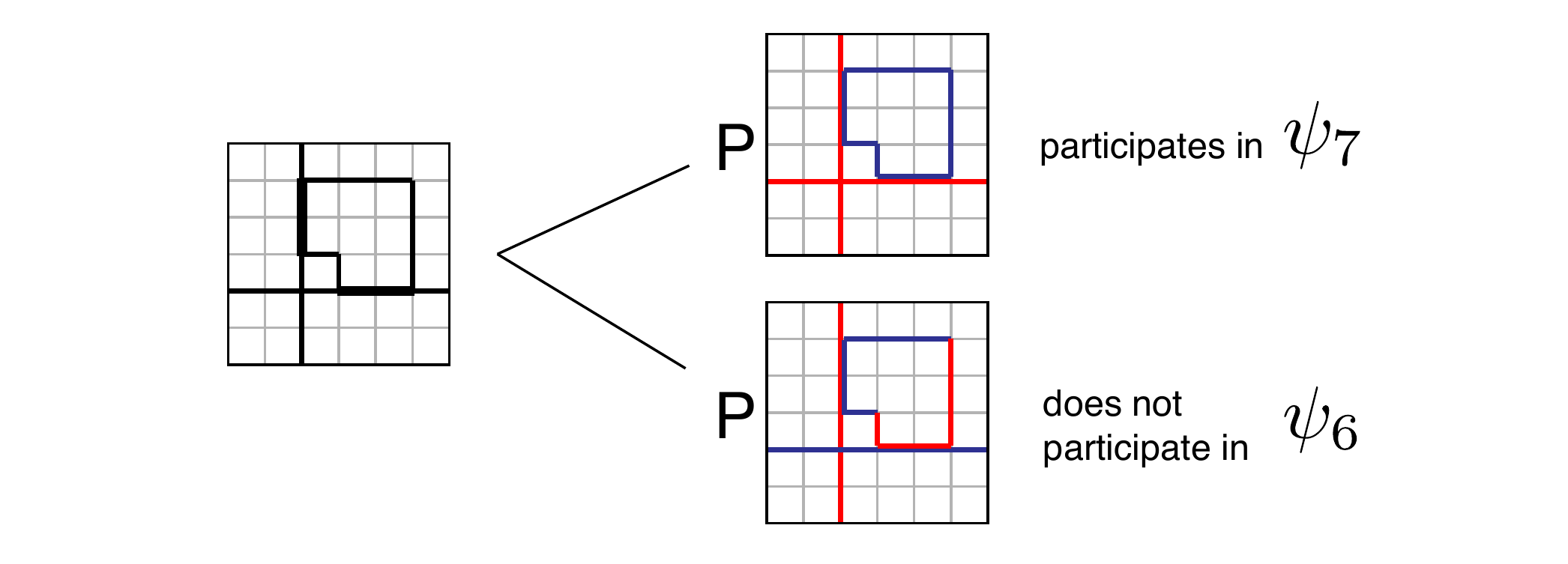}
\center
\caption{ 
The states $\Psi_6$ and $\Psi_7$ are two different ground states of the the spin-1 model on the torus, both involving two non-contractible loops. A loop configuration like the one in the figure, which includes a closed loop having common segments with both non-contractible loops, participates in the state $\Psi_7$, whereas it does not participate in the state
$\Psi_6$. In order to see this, we note that such configuration can result from merging two ground state configurations with both non-contractible loops belonging to the same copy (same color). It can not result, however, from merging two ground state configurations with the non-contractible loops belonging to different copies (different color). 
}
\label{DegTorusPru}
\end{figure}

For the spin-1 model we then have three degenerate ground states (Fig.\,\ref{DegAnnulus}), corresponding to the distinct states that result from merging the two possible Kitaev ground states:
\begin{eqnarray}
	\ket{\Psi_{1}} &=&
		\mathbf{P}\left[\ket{\Phi}\otimes\ket{\Phi}\right] \nonumber\\
\ket{\Psi_{2}} &=&
	\mathbf{P}\left[\mathcal{T}\ket{\Phi}\otimes\ket{\Phi}\right]= \mathbf{P}\left[\ket{\Phi}\otimes\mathcal{T}\ket{\Phi}\right]\nonumber \\
\ket{\Psi_{3}} &=&
\mathbf{P}\left[\mathcal{T}\ket{\Phi}\otimes\mathcal{T}\ket{\Phi}\right].	
\end{eqnarray}
Similarly, on the torus, there are $4$ degenerate ground states for Kitaev's model: $\ket{\Phi}$, $\mathcal{T}_x\ket{\Phi}$, $\mathcal{T}_y\ket{\Phi}$, and $\mathcal{T}_x\mathcal{T}_y\ket{\Phi}$, where $\mathcal{T}_{x(y)}=\prod_{\ell \in \mathcal{C}_{x(y)}} \sigma_{\ell}^x$, and $\mathcal{C}_{x(y)}$ are the two classes of non-contractible loops on the torus. After symmetrization, they give rise to $10$ degenerate ground states for the spin-$1$ model (Fig.\,\ref{DegTorus}). When counting the number of degenerate ground states it is important to note that the states 
\begin{eqnarray}
	\ket{\Psi_{6}} &=&
		\mathbf{P}\left[\mathcal{T}_x\ket{\Phi}\otimes \mathcal{T}_y \ket{\Phi}\right] \nonumber\\
	\ket{\Psi_{7}} &=&
		\mathbf{P}\left[\mathcal{T}_x\mathcal{T}_y\ket{\Phi}\otimes \,\ket{\Phi}\right], \nonumber
\end{eqnarray}
are different. They both involve two non-contractible loops, but they differ in whether these loops belonged to the same or different copies before merging. As a consequence of their different origin, the two states involve different loop configurations (see Fig.\,\ref{DegTorusPru}).


\section{Excitations} 
The toric code model exhibits two type of excitations, which correspond to quasiparticles localized either at vertices or plaquettes.
Vertex (plaquette) excitations are created by opening loops (dual loops) in the quantum loop condensate $\ket{\Phi}$ in the form:
\begin{eqnarray}
\ket{\Phi_{\eta_1\eta_2}}=\!\!	\prod_{\ell \in \Gamma_{\eta_1\!\eta_2}} 
	\!\!\!\!\sigma_{\ell}^x 
	\ket{\Phi},	 
	&& \ket{\Phi_{\xi_1\xi_2}}=\!\!\prod_{\ell \in \Gamma_{\xi_1\!\xi_2}} 
		\!\!\!\!\sigma_{\ell}^z 
		\ket{\Phi}.
		\label{KitaevExc}
\end{eqnarray}
Here, $\Gamma_{\eta_1\!\eta_2}$($\Gamma_{\xi_1\!\xi_2}$) is an open path with open ends located at vertices (plaquettes) $v=\eta_1,\eta_2$ $(p=\xi_1\xi_2)$. The state $\ket{\Phi_{\eta_1\eta_2}}$ ($\ket{\Phi_{\xi_1\xi_2}}$) is an eigenstate with eigenvalue $-1$ of the stabilizers $A_v(B_p)$  for $v=\eta_1,\eta_2$ $(p=\xi_1\xi_2)$.

The spin-1 model in Eq.\,(\ref{Hamiltonian}) displays gapped excitations localized at vertices or plaquettes, which result from merging vertex or plaquette excitations in the toric code copies. I do not know how to show that this Hamiltonian has a gap, but it seems very plausible. In such a case, these merged excitations are the elementary excitations of the system.

\subsection{Vertex excitations}

Let me start by creating two vertex excitations in the form:
\begin{eqnarray}
\ket{\Psi_{\eta_1\eta_2}}=
\mathbf{P}\left(\ket{\Phi_{\eta_1\eta_2}}\otimes\ket{\Phi}\right)
=\mathbf{P}(\ket{\Phi}\otimes\ket{\Phi_{\eta_1\eta_2}}).
\label{Spin1Vertex}
\end{eqnarray}
\begin{figure}[t!]
\includegraphics[width=\linewidth]{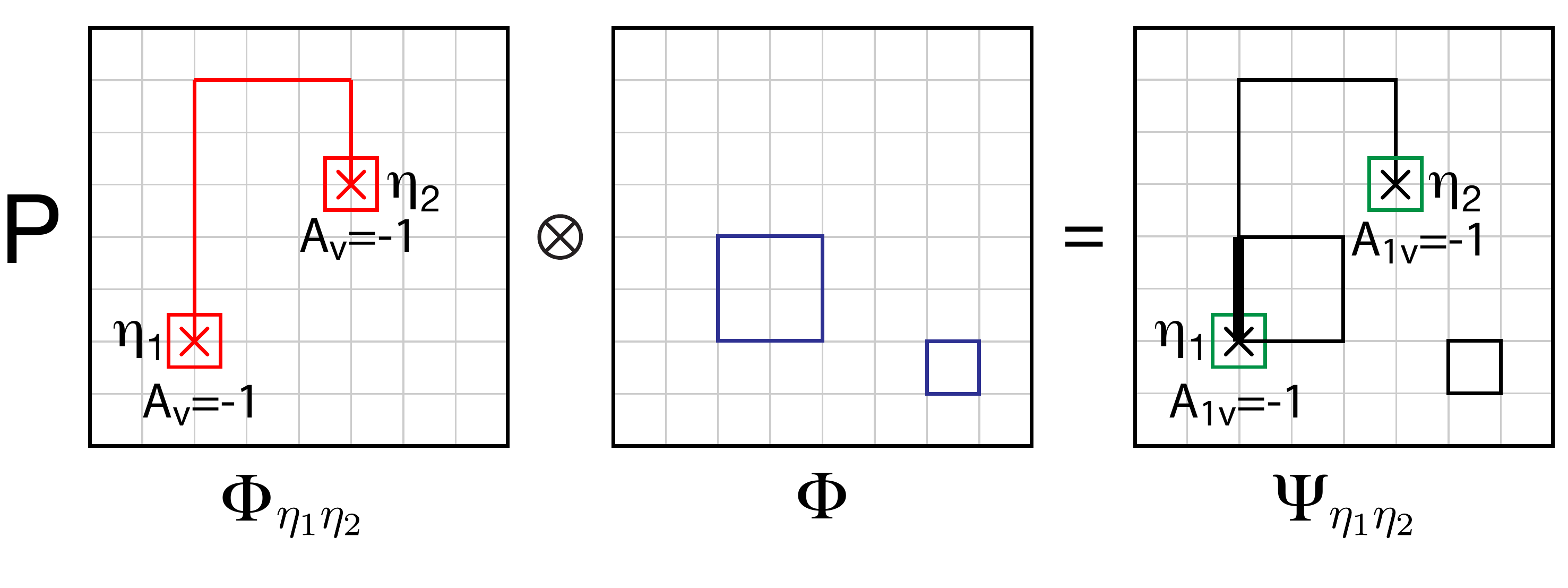}
\caption{\textbf{Localized vertex excitations of the spin-1 model}. Vertex excitations in a toric code copy, $\Phi_{\eta_1\eta_2}$, are localized at the ends of open loops. They are characterized by having eigenvalue $-1$ under the stabilizer $A_v$. Merging such excited state with a toric code ground state, $\Phi$, results in an excited state for the spin-1 model, $\Psi_{\eta_1\eta_2}$, with two excitations localized at the same vertices and characterized by having eigenvalue $-1$ under the operator $A_{1v}$.} 
\label{VertexExcitations}
\end{figure}
This state is an eigenstate of the Hamiltonian in Eq.\,(\ref{Hamiltonian}) with eigenvalue $2J$.  
It fulfills all plaquette conditions in Eq.\,(\ref{PlaquetteConditions}), but violates the first vertex condition in Eq.\,(\ref{VertexConditions}) for vertices $v=\eta_1,\eta_2$:
\begin{eqnarray}
A_{1v}\ket{\Phi_{\eta_1\eta_2}}=-\ket{\Phi_{\eta_1\eta_2}} && v= \eta_1,\eta_2.
\end{eqnarray}
These properties follow from the properties of the merged toric code states. For all plaquettes $p$ and for all vertices $v\neq \eta_1,\eta_2$, the toric code vertex excitation $\ket{\Phi_{\eta_1\eta_2}}$ fulfills the stabilizer conditions $B_p\ket{\Phi_{\eta_1\eta_2}}=A_v\ket{\Phi_{\eta_1\eta_2}}=\ket{\Phi_{\eta_1\eta_2}}$. Therefore the state in Eq.\,(\ref{Spin1Vertex}) satisfies the ground-state conditions in Eqs.\,(\ref{VertexConditions}) and (\ref{PlaquetteConditions}) and we have  
$Q_{iv(p)}\ket{\Psi_{\eta_1\eta_2}}=0$.
At vertices $v= \eta_1,\eta_2$, however, the excited toric copy fulfills $A_v\ket{\Phi_{\eta_1\eta_2}}=-\ket{\Phi_{\eta_1\eta_2}}$, and the state in Eq.\,(\ref{Spin1Vertex}) thus violates the first vertex condition:
\begin{eqnarray}
&& A_{1v}\ket{\Psi_{\eta_1\eta_2}}= \nonumber\\
&& =\mathbf{P}(A_v \otimes A_v) \ket{\Phi_{\eta_1\eta_2}}\otimes\ket{\Phi} 
=-\ket{\Psi_{\eta_1\eta_2}}.
\end{eqnarray}
The second vertex condition is not violated, since we have:
\begin{eqnarray}
A_{2v}\ket{\Psi_{\eta_1\eta_2}}
=A_{2v}A_{1v}\ket{\Psi_{\eta_1\eta_2}}
=-A_{2v}\ket{\Psi_{\eta_1\eta_2}}=0. 
\end{eqnarray}
Thus, the spin-1 state in Eq.\,(\ref{Spin1Vertex}) is an exact eigenstate of the spin-1 Hamiltonian, with two localized excitations at vertices $v= \eta_1,\eta_2$ characterized by  $Q_{1v}\ket{\Psi_{\eta_1\eta_2}}=\ket{\Psi_{\eta_1\eta_2}}$.

\begin{figure}[t!]
\includegraphics[width=\linewidth]{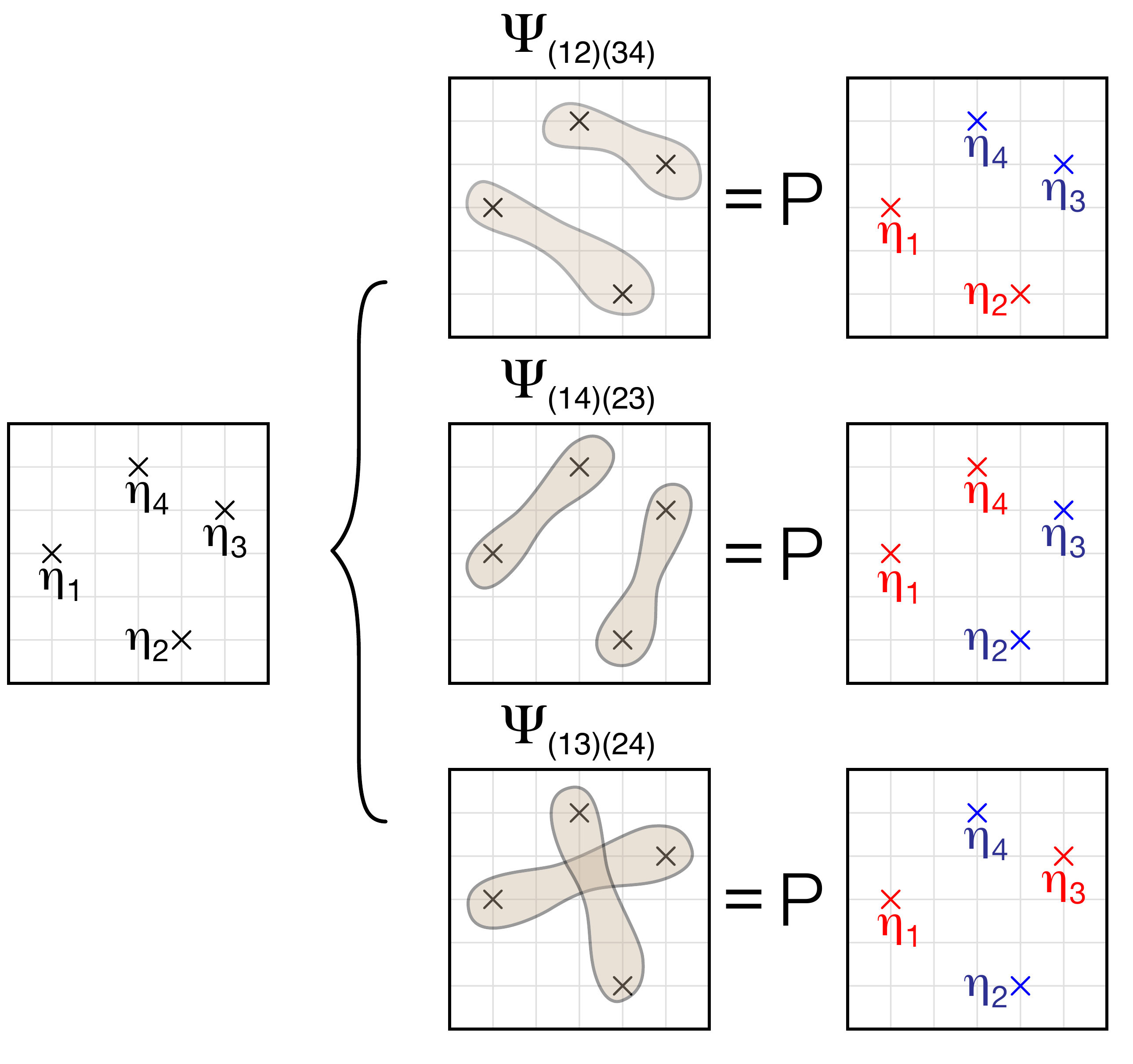}
\center
\caption{\textbf{Topological degeneracy of the subspace of 4 vertex excitations}. For fixed positions of four vertex quasiparticles, the different assignments of pairs to each of the indistinguishable copies give rise to three independent states which are locally indistinguishable.} 
\label{4VertexExcitations}
\end{figure}

\begin{figure}[ht]
\includegraphics[width=\linewidth]{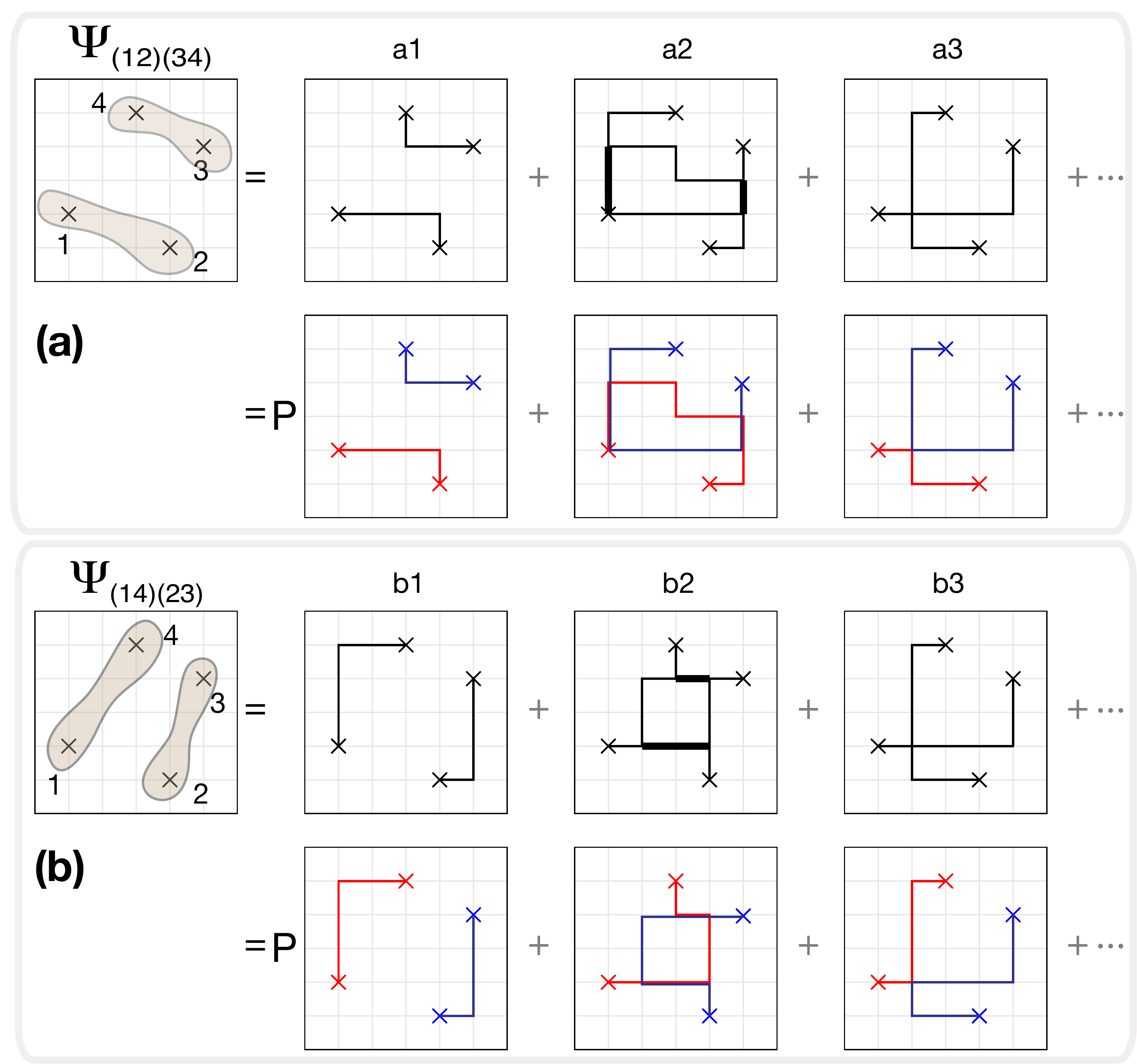}
\center
\caption{
The states \textbf{a,} $\Psi_{(1 2)(3 4)}$,   and \textbf{b,} $\Psi_{(1 4)(2 3)}$ result, respectively, from merging two toric code states in which quasiparticles $(1,2)$, $(3,4)$, and $(1,4)$, $(2,3)$ belong to the same copy. The state $\Psi_{(1 2)(3 4)}$ involves configurations of type (a1) in which quasiparticles $(1,2)$, and $(3,4)$ are attached by single-line segments of loop. These states do not participate in the state $\Psi_{(1 4)(2 3)}$. Similarly, configurations of type (a2), which can be decomposed into open loops joining $(1,4)$ and $(2,3)$ on one side, and a closed loop that has common segments with both open loops, on the other side, do not participate in the state $\Psi_{(1 4)(2 3)}$. The states $\Psi_{(1 2)(3 4)}$ and $\Psi_{(1 4)(2 3)}$ are not orthogonal, since configurations of type (a3, b3) appear in both.
}
\label{NonEquivalence}
\end{figure}

Following the structure of the state in Eq.\,(\ref{Spin1Vertex}) we can construct eigenstates with $2n$ vertex excitations, provided that they are located at $2n$ different vertices. Let us consider the case of $4$ vertex excitations (Fig.\,\ref{4VertexExcitations}). If the positions of the quasiparticles are fixed at vertices $\eta_1,\eta_2,\eta_3$ and $\eta_4$, we have still freedom to assign them to each of the toric code copies in one of the three forms $\ket{\Psi_{(\eta_1\eta_2)(\eta_3\eta_4)}}$, $\ket{\Psi_{(\eta_1\eta_4)(\eta_2\eta_3)}}$ or $\ket{\Psi_{(\eta_1\eta_3)(\eta_2\eta_4)}}$, where
\begin{eqnarray}
\ket{\Psi_{(\eta_1\eta_2)(\eta_3\eta_4)}}=
\mathbf{P}\left[
\ket{\Phi_{\eta_1\eta_2}}\otimes\ket{\Phi_{\eta_3\eta_4}}
\right].
\end{eqnarray}
States corresponding to different assignments are distinct. As shown in Fig.\,\ref{NonEquivalence}, the state $\ket{\Psi_{(\eta_1\eta_2)(\eta_3\eta_4)}}$ involves a certain class of loop configurations  which can not result from merging a loop with open ends at $\eta_1,\eta_4$ together with a one with open ends at $\eta_2,\eta_3$. Such configurations do not participate in the state $\ket{\Psi_{(\eta_1\eta_4)(\eta_2\eta_3)}}$. The different assignments can not be distinguished locally. They are all characterized by having eigenvalue $-1$ under the vertex operator $A_{1v}$ at vertices $v=\eta_1,\eta_2,\eta_3,\eta_4$. 

\begin{figure}[t!]
\includegraphics[width=\linewidth]{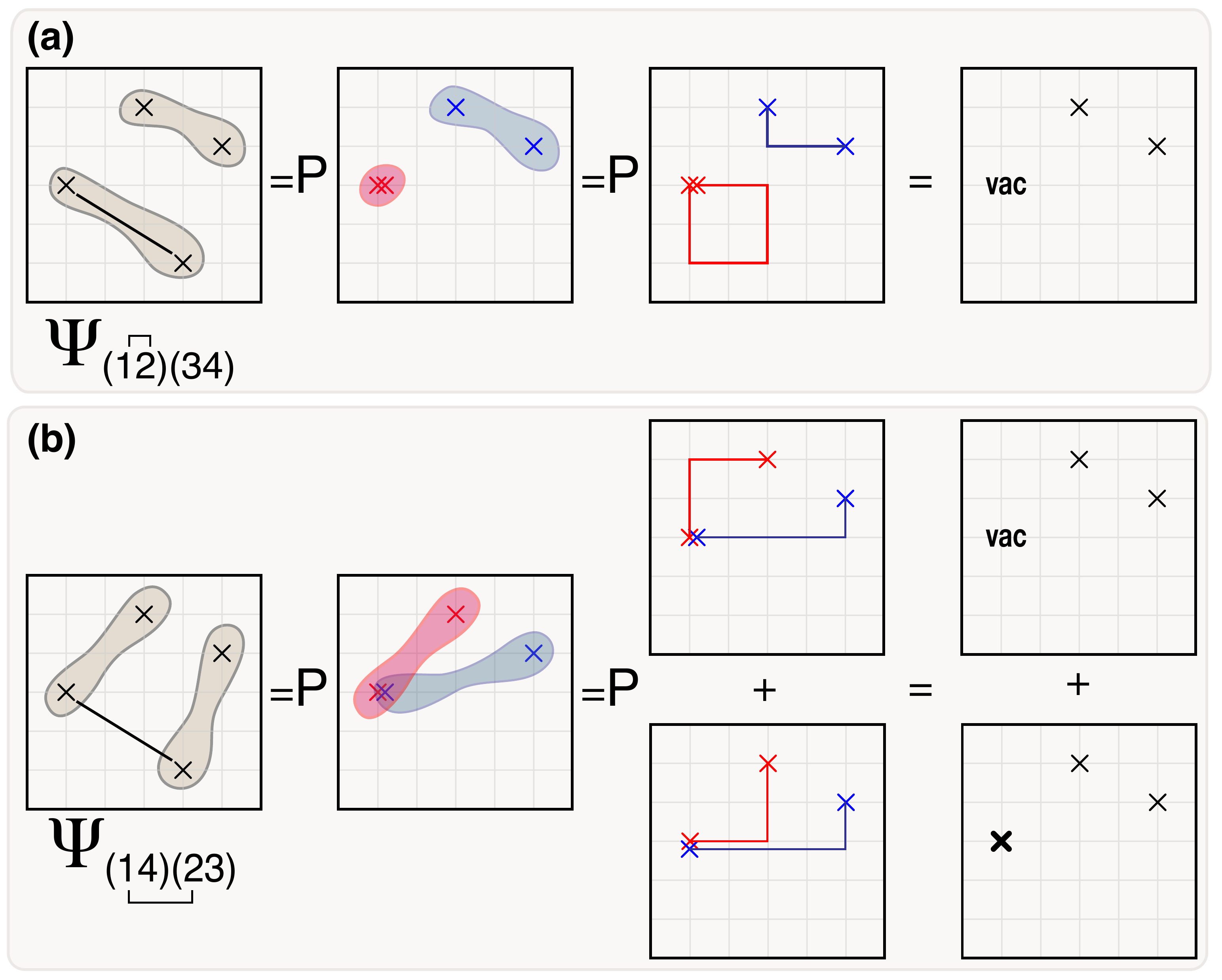}
\center
\caption{\textbf{Two Fusion Channels}. Vertex excitations fuse to the vacuum when they are assigned to the same copy. They can fuse to a vertex excitation if they come from different copies.
}
\label{FusionChannels}
\end{figure}

The different topological character of three degenerate 4-vertex states is manifested when bringing together two vertex quasiparticles (Fig.\,\ref{FusionChannels}). Let us fuse, for example, quasiparticles at vertices $\eta_1$ and $\eta_2$.  The result is different depending on whether the quasiparticles are or not assigned to the same copy:
\begin{eqnarray}
&&\ket{\Psi_{(\eta_1\eta_2)(\eta_3\eta_4)}} \xrightarrow[\eta_2\rightarrow \eta_1]{}
\ket{\Psi_{\eta_3\eta_4}}\nonumber\\
&&\ket{\Psi_{(\eta_1\eta_4)(\eta_2\eta_3)}} \xrightarrow[\eta_2\rightarrow \eta_1]{}
\ket{\Psi_{(\eta_1\eta_4)(\eta_1\eta_3)}}.
\end{eqnarray}
If they belong to the same copy, fusion gives rise to a not excited vertex at $\eta_1=\eta_2$. As in the toric code model, the quasiparticles fuse to the vacuum. In contrast, if quasiparticles come from different copies, fusion can lead either to a not excited vertex, or to an excited vertex
violating the second vertex condition in Eq.\,(\ref{VertexConditions}). 
Local measurement of the projector $Q_{2v}$ at the fused vertex, can therefore characterize (after fusion) the three degenerate states
\footnote{The state $\ket{\Psi_{(\eta_1\eta_4)(\eta_1\eta_3)}}$, resulting from the fusion of two vertex excitations belonging to different copies, is not an eigenstate of the Hamiltonian. It involves two type of configurations, those for which the vertex $\eta_1$ is not excited, and those for which it violates the second vertex condition. The latter have eigenvalue $1$ under the projector $Q_{2v}$, revealing the fact that the merged quasiparticles were coming from a different copy.}.

\subsection{Braiding plaquette and vertex excitations}

Plaquette excitations for the spin-1 model can be constructed similarly to vertex excitations above, in the form $\ket{\Psi_{\xi_1\xi_2}}=
\mathbf{P}\left(\ket{\Phi_{\xi_1\xi_2}}\otimes\ket{\Phi}\right)$. These excitations are characterized by violating the first plaquette condition in Eq.\,(\ref{PlaquetteConditions}) at plaquettes $p=\xi_1,\xi_2$:
\begin{eqnarray}
B_{1p}\ket{\Phi_{\xi_1\xi_2}}=-\ket{\Phi_{\xi_1\xi_2}} && p= \xi_1,\xi_2.
\end{eqnarray}

\begin{figure}[t]
\includegraphics[width=\linewidth]{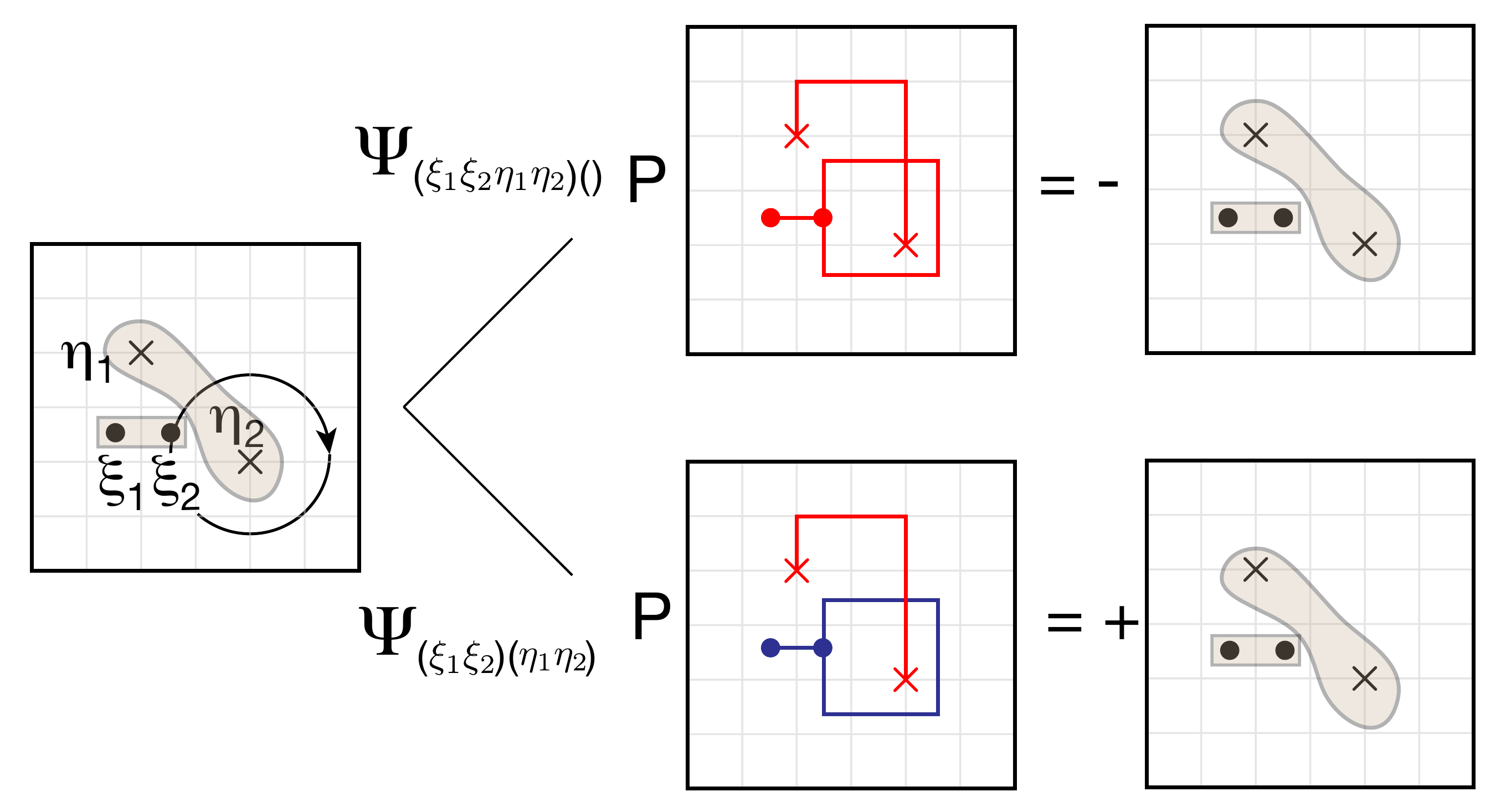}
\center
\caption{\textbf{Braiding of vertex excitations around flux excitations}. Braiding a vertex excitation around a flux excitation yields a phase $\pi$ if they are both assigned to the same copy, or a trivial phase if they are not.}
\label{Braiding}
\end{figure}

Let me consider the case in which a pair of plaquette excitations coexists with a pair of vertex excitations. There are two degenerate states:
\begin{eqnarray}
\ket{\Psi_{(\xi_1\xi_2\eta_1\eta_2)()}} & = &\mathbf{P}\left(\ket{\Phi_{\xi_1\xi_2\eta_1\eta_2}}\otimes\ket{\Phi}\right)\nonumber \\
\ket{\Psi_{(\xi_1\xi_2)(\eta_1\eta_2)}} & = &\mathbf{P}\left(\ket{\Phi_{\xi_1\xi_2}}\otimes\ket{\Phi_{\eta_1\eta_2}}\right),
\end{eqnarray}
depending on whether the two pairs belong to the same or different copies. These states are locally indistinguishable. They are characterized by the same eigenvalues of the projectors $Q_{1p}$ and $Q_{1v}$ (eigenvalue $1$ for $p=\xi_1,\xi_2$, $v=\eta_1,\eta_2$, and eigenvalue $0$ for the rest), which commute with each other. The two states can be distinguished by braiding plaquette and vertex excitations around each other (Fig.\,\ref{Braiding}). Braiding one plaquette excitation around a vertex excitation yields a phase $\pi$ if the two pairs belong to the same copy, whereas the phase is trivial if they do not: 
\begin{eqnarray}
\ket{\Psi_{(\xi_1\xi_2\eta_1\eta_2)()}} &\rightarrow&
-\ket{\Psi_{(\xi_1\xi_2\eta_1\eta_2)()}}\nonumber\\
\ket{\Psi_{(\xi_1\xi_2)(\eta_1\eta_2)}}  &\rightarrow&
\ket{\Psi_{(\xi_1\xi_2)(\eta_1\eta_2)}}.
\end{eqnarray}
The braiding is therefore described by a $2\times2$ matrix, which performs a rotation in the quasiparticle topological subspace. These non-trivial braiding properties imply non-trivial fusion properties. Plaquette and vertex excitations have two fusion channels: they can fuse to a fermion (if they belong to the same copy) or to the vacuum (if they do not).

\section{Conclusions and outlook} 
I have presented a class of states that result from merging identical quantum loop condensates. I have proposed that these states are good candidates to describe non-Abelian topological phases. This possibility has been illustrated through a spin-$1$ model, that results from combining two toric code Abelian theories.  The resulting model is a sum of vertex and plaquette projectors that are not mutually commuting. It seems very plausible that this Hamiltonian has a gap. I believe that proving the existence of a gap might be not straightforward, as it happens for loop models with non-orthogonal inner products \cite{Fendley2008}, for which the existence of a gap has not been proven.
The dependence of the ground state degeneracy on the topology of the lattice confirms that the model is topological. 
Moreover, the non-trivial braiding and fusion properties of the gapped excitations I described, strongly indicate that the model is non-Abelian. This is also supported by the fact that the ground-state degeneracy on the torus is not a perfect square, as one would expect for a non-chiral Abelian model. It is challenging to identify the particular anyon model to which this model corresponds. It is interesting to note that the $3 \times 3$ matrix $U$, performing the unitary transformation to the dual representation, coincides with the $S$-matrix characterizing Ising anyons \cite{Preskill}. The $10$-fold degeneracy on the torus does not match, however,  the one of a doubled Ising topological quantum field theory, which should be $9$-fold degenerate. Whether or not the present model is related to an Ising anyon model is unclear to me at this point.

The merging construction can be generalized to an arbitrary number of copies $k$ and to arbitrary type of Abelian theories to be merged. For the case of $k$ toric-code copies, the resulting spin-$\frac{k}{2}$ models are characterized by the unitary matrix $U^{(k)}=P^{(k)}\mathsf{H}^{\otimes k}P^{(k)}$, where $P^{(k)}$ is the local projector onto the symmetric subspace of the $\frac{k}{2}$ spin, and $\mathsf{H}$ is the Hadamard transformation. The matrix $U^{(k)}$ describes the change of basis to the dual lattice representation. It seems plausible that these models might display $SU(2)_k$ anyons.  It is then interesting to consider the case of spins $\frac{3}{2}$, which might allow for universal fault-tolerant quantum computation \cite{Preskill}. 

The spin-$1$ model presented here could be realized in experiments with ultracold atoms or molecules in optical lattices, for which theoretical protocols have been designed to achieve four-spin interactions around plaquettes or vertices \cite{Zoller}. It is challenging to design a theoretical protocol in which the excitations described here could be created and manipulated to test their non-trivial braiding and fusion properties.

The construction I have described might be equivalent to string-net or generalized loop formalisms proposed in the literature. %
Within the merging picture introduced here, 1) the non-Abelian state is intuitively constructed by combining two Abelian ground states, 2) the non-Abelian braiding properties arise in a transparent way as the consequence of  the indistinguishability of the Abelian states that are merged, and 3) the parent Hamiltonian is simple, involving four spin interactions with vertex and plaquette operators related by a unitary transformation, which seems to characterize the anyon model.
  
Moreover, the merging picture could be applied to describe {\em chiral} non-Abelian phases, like those expected to occur in fractional quantum Hall systems \cite{Stern2010}, for which it is unclear whether a string-net condensation picture is appropriate \cite{Levin2005}. Indeed, the celebrated Pfaffian state \cite{MooreRead91} can be written as the result of merging two (Abelian) Laughlin states \cite{Paredes2012}. This suggests the possibility of a unified picture, in which a non-Abelian phase would arise from the combination of two or more Abelian ones.
\newline

I would like to thank Miguel Aguado and Xiao-Gang Wen for helpful comments on this manuscript.

\appendix

\section{APPENDIX: DETAILS OF PROOF OF UNIQUENESS OF GROUND STATE}

\subsection{A. Characterization of ground state subspace}
Graphs that can not be decomposed into two sets of closed loops do not participate in the ground state of the Hamiltonian in Eq.\,(\ref{Hamiltonian}). 

\begin{figure}[t!]
\includegraphics[width=0.9\linewidth]{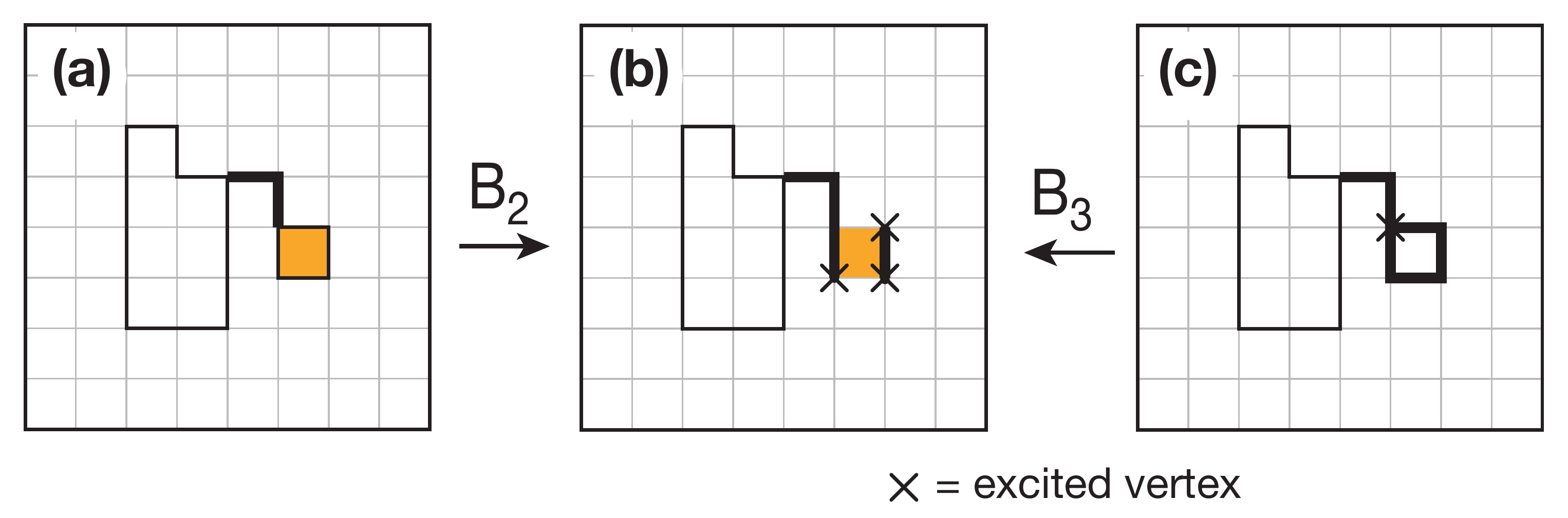}
\center
\caption{
A configuration of the form  \textbf{(a)} can not participate in the ground state superposition. For the selected plaquette $p$, a $B_{2p}$-move leads to the configuration \textbf{(b)}, with three excited vertices in that plaquette. Such configuration can not be cancelled by performing a $B_{3p}$-move on any configuration with no-vertex excitations. Configuration \textbf{(b)} could be cancelled by a $B_{3p}$-move acting on the configuration \textbf{(c)}, but the latter does have an excited vertex.
}
\label{ThreeVertex}
\end{figure}

Let us assume that a state of the form in Fig.\,\ref{TwoClassGraphs}b appears in the ground state superposition. Since this state is connected through a sequence of $B_{2p}$ moves to a state of the form in Fig.\,\ref{ThreeVertex}, it follows that the latter appears also in the superposition. In order to be a ground state, such superposition has to fulfill the second plaquette condition in Eq.\,(\ref{PlaquetteConditions}), which implies that any $B_{2p}$ move has to be cancelled by a $B_{3p}$ move.
But there is a $B_{2p}$ move that converts the state in Fig.\,\ref{ThreeVertex} into a state with three vertex excitations on the same plaquette. Such state can not be cancelled by a $B_{3p}$ move on the same plaquette. 

\subsection{B. Amplitudes of loop configurations}
Cancellation of $B_2$- and $B_3$-moves in a state of the form of Eq.\,(\ref{GenGroundState}) uniquely determines the coefficients to be $\beta_\mathcal{L} \propto 2^{n_{\mathcal{L}}}$. 

Satisfying the second plaquette condition in Eq.\,(\ref{PlaquetteConditions}) implies that for every single-line loop configuration $\vert \mathcal{L}_s \rangle$, and every plaquette, we have $\langle \mathcal{L}_s\vert B_{2p}\vert \Psi \rangle =\langle \mathcal{L}_s\vert B_{3p} \vert \Psi \rangle$.  We distinguish between two possible different cases. 

\begin{figure}[t]
\includegraphics[width=0.8\linewidth]{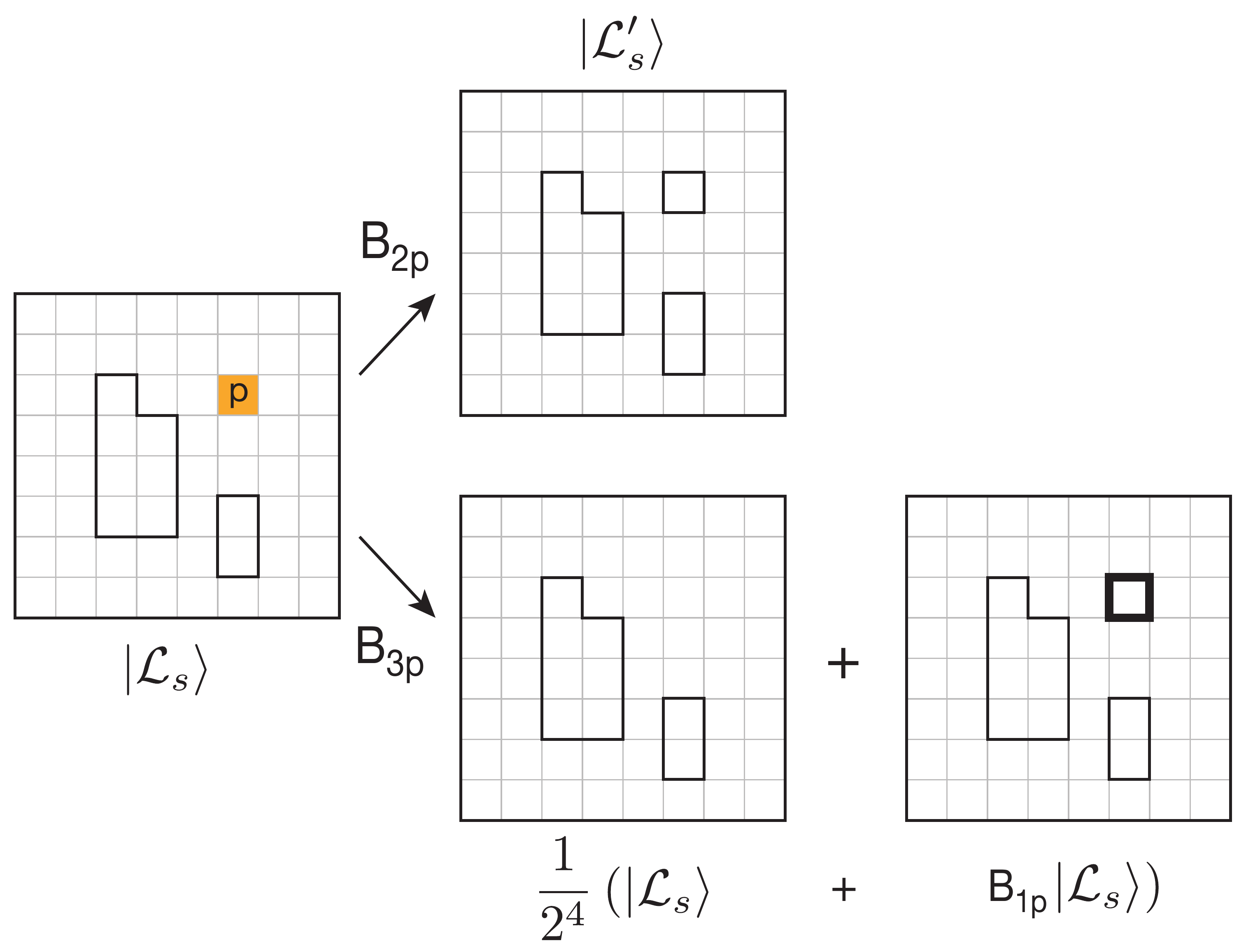}
\center
\caption{\textbf{Plaquette moves (case 1)}.
Plaquette-moves $B_{2p}$ and $B_{3p}$ acting on a configuration $\vert \mathcal{L}_s \rangle$, for the case in which the loop configuration has no common links with the plaquette $p$. The $B_{2p}$-move leads to a configuration $\vert \mathcal{L}_s^\prime \rangle$ with one more loop.
The $B_{3p}$-move leads to a superposition of $\vert \mathcal{L}_s \rangle$ and $B_{1p}\vert \mathcal{L}_s \rangle$.}
\label{PlaquetteMove1}
\end{figure}

1) The plaquete $p$ does not have links in common with the loop configuration $\mathcal{L}$, and therefore the $B_{2p}$-move converts the state $\vert \mathcal{L}_s \rangle$ into a configuration $\vert \mathcal{L}^\prime_s \rangle$ with one more single-line loop ( Fig.\,\ref{PlaquetteMove1}). For this case we have that 
$\langle \mathcal{L}_s^\prime\vert \mathcal{L}_s^\prime \rangle =2^{-4} \langle \mathcal{L}_s\vert \mathcal{L}_s \rangle$, and that
\begin{eqnarray}
\langle \Psi \vert B_{2p}\vert \mathcal{L}_s \rangle  &=&  \langle \Psi \vert \mathcal{L}^\prime_s \rangle=  
\beta_\mathcal{L^\prime}\langle \mathcal{L}^\prime_s\vert \mathcal{L}^\prime_s \rangle \nonumber \\
\langle \Psi \vert B_{3p}\vert \mathcal{L}_s \rangle &=&  \frac{1}{2^{4}} \langle \Psi \vert \left(1+B_{1p}\right)\vert \mathcal{L}_s \rangle=  
2 \beta_{\mathcal{L}}\langle \mathcal{L}^\prime_s\vert \mathcal{L}^\prime_s\rangle\nonumber.
\end{eqnarray}
Therefore we conclude that in this case $\beta_{\mathcal{L}^{\prime}}=2\beta_{\mathcal{L}^{}}$.

\begin{figure}[t]
\includegraphics[width=0.8\linewidth]{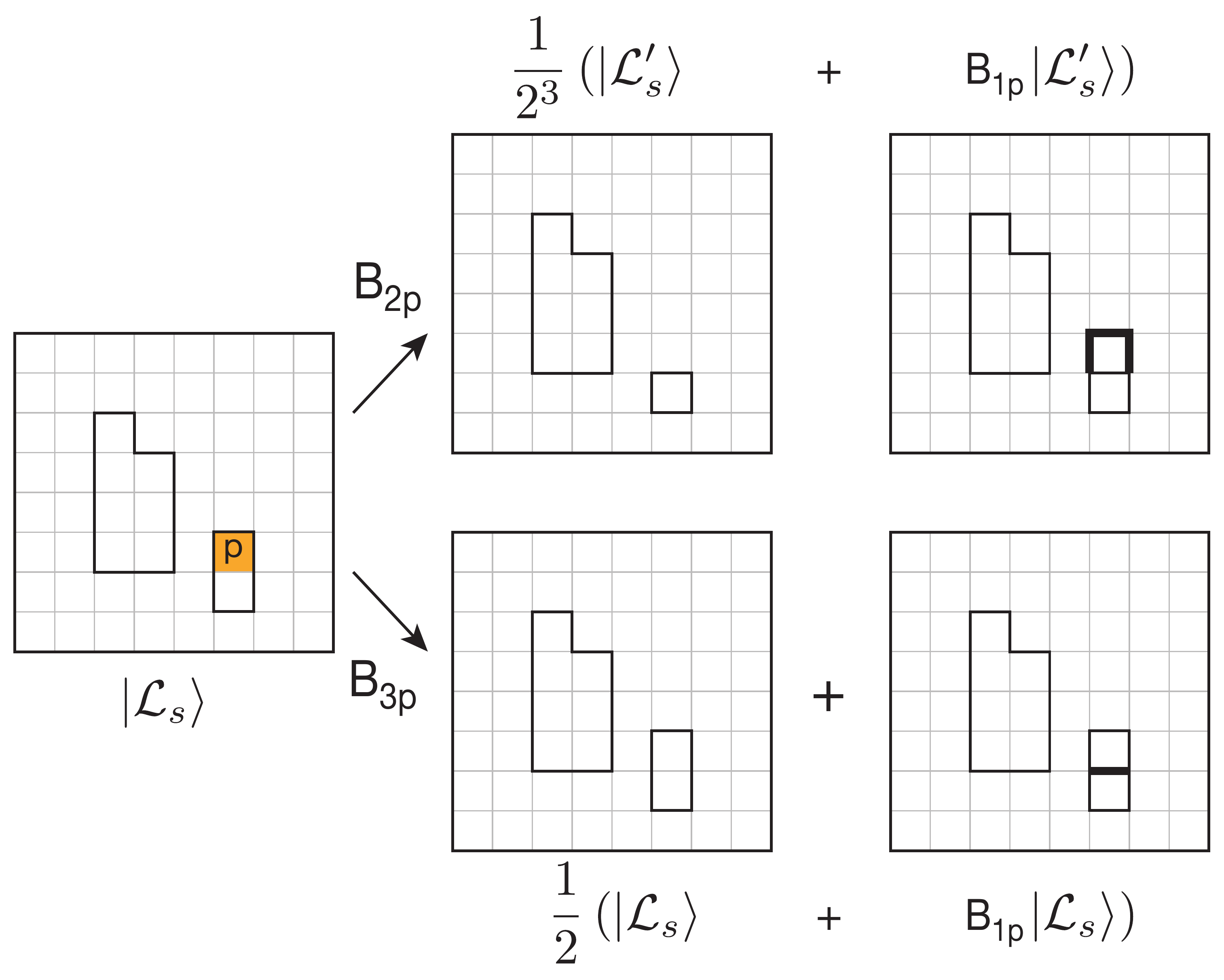}
\center
%
\caption{\textbf{Plaquette moves (case 2)}.
Plaquette-moves $B_{2p}$ and $B_{3p}$ acting on a configuration $\vert \mathcal{L}_s \rangle$, for the case in which the loop configuration has common links with the plaquette $p$. The $B_{2p}$-move leads to a configuration $\vert \mathcal{L}_s^\prime \rangle$ with the same number of loops. The $B_{3p}$-move leads to a superposition of $\vert \mathcal{L}_s \rangle$ and $B_{1p}\vert \mathcal{L}_s \rangle$.
The corresponding amplitudes depend on the number of common links of the loop configurations $\mathcal{L}$ and $\mathcal{L}^\prime$ with the plaquette $p$.}
\label{PlaquetteMove2}
\end{figure}

2) The plaquete $p$ does have links in common with the loop configuration $\mathcal{L}$, and therefore the $B_{2p}$-move gives rise to a configuration $\vert \mathcal{L}^\prime_s \rangle$ with the same number of single-line loops than $\vert \mathcal{L}_s \rangle$ ( Fig.\,\ref{PlaquetteMove2}). For this case we have that 
\begin{eqnarray}
\langle \Psi \vert B_{2p}\vert \mathcal{L}_s \rangle  &=&  	\frac{1}{2^{n_{1p}}} \langle \Psi \vert \left(1+B_{1p}\right)\vert \mathcal{L}^\prime_s \rangle=  
\frac{2}{2^{n_{1p}}} \beta_{\mathcal{L}^\prime}\langle \mathcal{L}^\prime_s\vert \mathcal{L}^\prime_s \rangle \nonumber\\
\langle \Psi \vert B_{3p}\vert \mathcal{L}_s \rangle &=&  \frac{1}{2^{n_{1p}^\prime}} 
\langle \Psi \vert \left(1+B_{1p}\right)\vert \mathcal{L}_s \rangle =
\frac{2}{2^{n_{1p}^\prime}} \beta_{\mathcal{L}}^{}\langle \mathcal{L}_s^{}\vert \mathcal{L}_s^{} \rangle, \nonumber
\end{eqnarray}
where $n_{1p}$ and $n_{1p}^\prime$ are, respectively, the number of common links of the plaquette $p$ with the loop configurations $\mathcal{L}$ and $\mathcal{L}^\prime$. Since we have that $\langle \mathcal{L}_s^\prime\vert \mathcal{L}_s^\prime \rangle =2^{n_{1p} -n_{1p}^\prime} \langle \mathcal{L}_s\vert \mathcal{L}_s \rangle$, we conclude that $\beta_{\mathcal{L}^{\prime}}=\beta_{\mathcal{L}^{}}$. 

Therefore, loop configurations with the same number of loops have equal amplitudes, whereas those differing by a closed loop have a relative amplitude two.

\end{document}